\documentclass[english,10pt,oneside,journal]{IEEEtran}
\usepackage{definitions}

\begin{document}
\title{Bitwise Retransmission Schemes for Resources Constrained Uplink Sensor
  Networks} \author{Mohamed~A.~M.~Hassanien, Pavel~Loskot, Salman M. Al-Shehri,
  Tolga Numano\v{g}lu, and Mehmet Mert \thanks{MAH, PL and SMA are with the
    College of Engineering, Swansea University, Bay Campus, Swansea SA1 8EN,
    United Kingdom (email: \{mohmmedali25,salman777881\}@hotmail.com,
    p.loskot@swan.ac.uk)} \thanks{TN and MM are with Aselsan A.S.,
    Communications and IT Division, Ankara, Turkey (email:
    \{tnumanoglu,mmert\}@aselsan.com.tr)}\thanks{Corresponding author: Pavel
    Loskot, tel.: +44 1792 602619, fax: +44 1792 295676}}
\maketitle

\begin{abstract} Novel bitwise retransmission schemes are devised which
  retransmit only the bits received with small reliability. The retransmissions
  are used to accumulate the reliabilities of individual bits. Unlike the
  conventional automatic repeat request (ARQ) schemes, the proposed scheme does
  not require a checksum for the error detection. The bits to be retransmitted
  are reported as a combination number, or two synchronized random number
  generators (RNGs) at the transmitter and receiver are used to greatly
  compress the feedback message. The bitwise retransmission decisions and/or
  combining can be performed after the demodulation or after the channel
  decoding at the receiver. The bit-error rate (BER) expressions are derived
  for the case of one and two retransmissions, and verified by computer
  simulations. Assuming three specific retransmission strategies, the scheme
  parameters are optimized to minimize the overall BER. For the same number of
  retransmissions and packet length, the proposed schemes always outperform the
  frequently used stop-and-wait ARQ. The impact of feedback errors is also
  considered. Finally, practical designs of the bitwise retransmissions for
  data fusion from sensor nodes in Zigbee, Wifi and Bluetooth networks are
  presented.
\end{abstract}


\IEEEpeerreviewmaketitle

\section{Introduction}

It is well-known that feedback cannot improve the information theoretic
capacity of memoryless channels \cite{Goldsmith97}. However, the availability
of feedback can greatly simplify the encoding and decoding complexity
\cite{Cover91}. A good example of such schemes with the reduced implementation
complexity due to feedback are the ARQ retransmission schemes
\cite{Proakis,Lin83}. The retransmission schemes are optimized to trade-off the
reliability (e.g., the BER), the throughput (or equivalently, delay) and the
implementation complexity \cite{Beh07} or the energy consumption \cite{Wang16}.
For instance, the retransmissions can comprise a smaller number of bits than in
the original packet. The incremental redundancy hybrid ARQ (IR-HARQ) or type-II
HARQ schemes progressively reduce the coding rate of the forward error
correction (FEC) code with each additional retransmission at the expense of
increasing the decoding delay and reducing the throughput \cite{Ji05}. The
retransmission decision delays in HARQ schemes were reduced in \cite{Pai11} by
exploiting the structure of tail-biting convolutional codes. The permutations
of bits in the retransmitted packets are used in \cite{Wengerter02} and
\cite{Jung06} to improve the reliability of ARQ schemes. A holistic design of
the complexity-constrained type-II hybrid ARQ schemes with turbo codes is
considered in \cite{Chen13}. The joint design of FEC coding for forward data
delivery and reverse feedback signaling is studied in \cite{Lucani12} and
\cite{Moreira15}. The IR diversity and the time-repetition (TR) diversity are
compared in \cite{Makki14}. The time and superposition-coding packet sharing
between two independent information flows is optimized in \cite{Jabi15}; the
latter is shown to have a slightly better performance at the expense of larger
design and implementation complexity. Such transmission sharing strategies can
significantly outperform the conventional HARQ schemes when signal-to-noise
ratio (SNR) is sufficiently large \cite{Jabi15}. Assuming Gaussian transmission
codebooks, the achievable throughput of the HARQ schemes are compared to the
ergodic channel capacity in \cite{Jabi15}, and to the delay-limited channel
capacity in \cite{Makki14}. The transmission powers of ARQ schemes are
optimized in \cite{Makki14}.

Since the received packets typically contain only a few transmission errors,
the retransmission efficiency can be improved by the partial ARQ schemes
\cite{Ahmed15}. A truncated type II hybrid ARQ over block fading channels is
considered in \cite{ESA}. In addition to ARQ schemes exploiting a variable
number of retransmissions, the variable rate ARQ schemes optimize the number of
bits in each retransmission \cite{Modiano99,Makki14,Ahmad15}. The number of
retransmission bits required for a successful decoding is estimated from the
mutual information in \cite{Ahmad15} and \cite{Jabi15}. The pre-defined
retransmission patterns are assumed in \cite{Ahmed15}. The reactive
rate-adaptive ARQ strategies \cite{Ji05} usually outperform the proactive
strategies \cite{Modiano99}. The FEC coding to recover from the transmission
errors may be preferred to the ARQ retransmissions in case of multimedia
\cite{Shih14}. Except the stop-and-wait ARQ schemes, the go-back-N ARQ
retransmissions are optimized, for example, in \cite{Fujii10}. A multi-bit
feedback signaling to improve the ARQ performance is considered in
\cite{Hassanien10} and \cite{Jabi15}. Other papers account for more realistic
design constraints such as a noisy feedback \cite{ESA,Makki14}, and the channel
estimation in \cite{Ahmed15}. It is shown in \cite{Makki14} that the errors of
1 and 2-bit feedback messages can be neglected if their bit error probability
is less than $10^{-3}$, or if these errors are compensated by a non-uniform
allocation of the transmission powers. Furthermore, the repetition diversity
appears to be more robust to feedback errors than the IR-HARQ \cite{Makki14}.

In our conference paper \cite{Hassanien10}, we assumed a multi-bit error-free
feedback to evaluate the performance trade-offs between the throughput and the
reliability of the segmentation-based and the bitwise ARQ retransmission
schemes. Both schemes were found to outperform the stop-and-wait ARQ with the
bitwise ARQ providing larger reliability gains than the segmentation-based ARQ,
albeit at the expense of a greater complexity of the feedback signaling. In
this paper, we revisit the selective bitwise retransmission scheme proposed in
\cite{Hassanien10} to carry out more rigorous performance analysis and to
optimize its design. Recall that the proposed scheme aims to accumulate the
reliabilities of the least reliable individual bits in the received packet, so
it does not require the cyclic redundancy check (CRC) bits to make the
retransmission decisions. It can be combined with other FEC coding schemes
where the bit reliability is evaluated either before or after the FEC decoding.
From the implementation point of view, it is useful to keep the number of
retransmitted bits as well as the number of retransmissions constant, for
example, to maintain a constant transmission delay and throughput for each data
packet. However, a variable rate scheme that retransmits all bits having their
reliability below a given threshold is also investigated. The bits with small
reliabilities are reported back to the sender using either a binomial
combination number, or using a deterministic sequence of bit-permutations. Our
analysis assumes multi-bit error free feedback signaling, however, we also
evaluate the conditions when such assumption is justifiable. Finally, we
consider a more specific system-level design of the proposed bitwise
retransmission scheme to be employed in an uplink data collection scenario from
the resources-constrained sensors to a data fusion access point assuming
time-sharing of the transmission channel among the network nodes. The resulting
transmission protocol creates fully occupied packets with the retransmitted
bits having a higher priority than the newly arrived information bits.

The rest of this paper is organized as follows. System model is introduced in
Section II. The mathematical theoretic analysis of the proposed bitwise
retransmission scheme is carried out in Section III. The retransmission
protocols are presented and optimized in Section IV. The uplink data fusion
bitwise ARQ scheme is designed in Section V. Conclusions are given in Section
VI.

\section{System Model}

The design and analysis of the proposed bitwise retransmission scheme assumes a
point-to-point duplex communication link between a source node and a
destination node. As shown in \cite{Makki14}, the feedback errors in ARQ
schemes can be neglected provided that their probability of error is
sufficiently small. Hence, in our analysis, we assume the error-free feedback,
and in Section IV, we estimate the acceptable probability of feedback bit
errors when such assumption is justified in practice.

Assuming that $M$-ary modulation is used for transmission from the source to
the destination node, prior to the FEC decoding, the reliability of the
received bit $b_i$ can be calculated as \cite{Big05},
\begin{equation}\label{eq:1}
  \Lambda(b_i|y)= \log \frac{\Prob{b_i=0|y}}{\Prob{b_i=1|y}},\
  i=1,2,\ldots,\log_2 M
\end{equation}
where $y$ is the received $M$-ary modulation symbol. For equally probable
modulation symbols, assuming coherent detection in a channel with additive
white Gaussian noise (AWGN), the reception of modulation symbols can be
mathematically modeled as,
\begin{equation*}
  y= h\,s+w
\end{equation*}
where $h$ is a residual attenuation, possibly after the (multipath)
equalization, and $w$ is a zero-mean sample of AWGN. The log-likelihood ratio
(LLR) \eref{eq:1} can be then rewritten as,
\begin{equation}\label{eq:2}
  \Lambda(b_i|y)= \log \frac{\sum_{s\in
      S(b_i=0)}\exp\left(-|y-hs|^2/N_0\right)}{\sum_{s\in S(b_i=1)} \exp\left(
      -|y-hs|^2/N_0\right)}
\end{equation}
where the $M$-ary modulation constellation is partitioned as,
$S=S(b_i=0)\cup S(b_i=1)$, depending on the value of the $i$-th bit $b_i$ in
modulation symbols, $N_0$ is the power spectral density of the AWGN, and
$|\cdot|$ denotes the absolute value. 

The LLR \eqref{eq:1} can be also used as the soft-input decisions to the FEC
decoder. On the other hand, we can also use the soft-output decisions produced
by the decoder to make the retransmission decisions. In this case, the sums in
\eqref{eq:2} are done over the corresponding codewords mapped to sequences of
$M$-ary modulation symbols, and the sums in \eref{eq:2} are often approximated
using the formula, $\log\sum_i \exp(-a_i) \approx \min(a_i)$.

In order to simplify the analysis and illustrate the main concepts, in the
sequel, we consider a binary antipodal modulation. Generalization to higher
order modulations is straightforward by scaling the demodulated binary symbols
by appropriate constants computed for a given $M$-ary modulation scheme. Hence,
assume the transmitted binary symbols $S_1=\sqrt{E_b}$ and $S_2=-\sqrt{E_b}$,
so the received $N$-bit packet can be written as,
\begin{equation*}
  r_i= S_{ji}+w_i,\ i=1,2,\ldots,N
\end{equation*}
where $S_{ji}\in\{S_1,S_2\}$, and the zero-mean AWGN samples $w_i$ have the
constant variance, $\E{|w_i|^2}= \sigma_w^2= N_0/(2|h|^2)$. Then, the
reliabilities of the received bits are \cite{Big05},
\begin{equation*}
  \Lambda(b_i|r_i)\propto |r_i/\sigma_w|\equiv |\rb_i|.
\end{equation*}

Finally, the SNR per binary modulation symbol is defined as $\gammab= \Eb/N_0$.
If $\Rf$ denotes a fraction of information bits in the sequence of all bits
transmitted from the source to the destination, for a fair comparison of
different schemes with various rates $\Rf$, we assume the SNR per bit,
$\gammab= \Eb\Rf/N_0$.

\subsection{Bitwise retransmissions}

\begin{figure}[!t]
  \centering
  \includegraphics[scale=0.55]{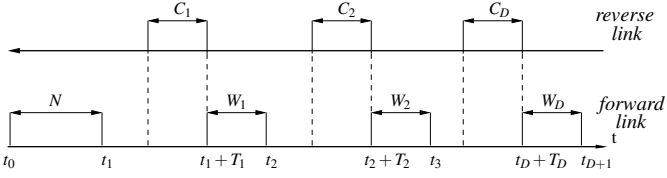}
  \caption{The timings of the proposed bitwise retransmission protocol.}
  \label{Fig:02}
\end{figure}

After the initial transmission of $N$-bit data packet, the destination uses the
received bit reliabilities to decide which $W_d$ binary symbols among $N$
received, $0\leq W_d \leq N$, should be retransmitted, where $d=1,2,\ldots,D$
denotes the retransmission index. The retransmission request is a feedback
message of $C_d\geq 1$ bits sent to the source node from the destination node
via a reverse (feedback) link. Consequently, after $D$ retransmissions, the
total number of bits sent over the forward link to the destination is,
\begin{equation*}
  \Nf=N+\sum_{d=1}^D W_d
\end{equation*}
whereas the total number of bits sent from the destination to the source over
the reverse link is,
\begin{equation*}
  \Nr=\sum_{d=1}^D C_d.
\end{equation*}
For instance, the conventional stop-and-wait ARQ scheme has the parameters
$W_d=N$ and $C_d=1$ (ACK or NACK). The retransmitted bits are combined using a
maximum ratio (MRC) or other combining method to improve their reliability. The
timings of the bitwise retransmission protocol are shown in \fref{Fig:02} where
$T_d$ represents the delay until the start of transmission of the window of
$W_d$ retransmitted bits.
The forward transmission rate is then defined as,
\begin{equation}\label{eq:10}
  \Rf(D)= \frac{N}{N+\sum_{d=1}^D W_d}= \frac{N}{\Nf}
\end{equation}
and the reverse transmission rate is defined as,
\begin{equation*}
  \Rr(D)= \frac{\sum_{d=1}^D C_d}{N+\sum_{d=1}^D C_d}= \frac{\Nr}{N+\Nr}.
\end{equation*}
Note that the feedback delays $T_d$ are not included in the definition of these
rates, since during these time intervals, both forward and reverse links are
possibly available for other transmissions.

In order to simplify the notation, let $W_1=W_2=\ldots=W_D=W$, and
$C_1=C_2= \ldots=C_D=C$, and we drop the bit index within the packet unless
stated otherwise. The $W$ least reliable bits requested for the next
retransmission can be identified by sorting the received bits by their
reliabilities. A typical sorting algorithm has the complexity $O(N\log N)$
\cite{Knuth98}. The locations of these $W$ bits within the packet of $N$ bits
can be reported back to the sender by the corresponding binomial number. The
number of bits representing such feedback message is,
\begin{equation*}
  C= \left\lceil \log_2 \binom{N}{W} \right\rceil
\end{equation*}
where $\binom{N}{W}$ is the binomial number, and $\lceil\cdot\rceil$ is the
ceiling function. More importantly, even though typically $C>W$, we always have
that, $\Rf\gg\Rr$, as in many other ARQ protocols. Thus, the multi-bit feedback
can be very beneficial to improve the performance of ARQ protocols
\cite{Hassanien10}.

\section{Performance Analysis}

Given the values of $N$, $W$, $C$, $D$ and $\gammab$, we now derive the average
BER of the proposed bitwise retransmission scheme. We first assume the case of
a single retransmission, i.e., $D=1$. We optimize the number of feedback
message bits $C$ the forward throughput $\Rf$ to minimize the BER, and compare
our scheme to the conventional stop-and-wait ARQ. We then perform the similar
analysis for $D=2$ retransmissions before generalizing the obtained BER
expressions to the case of $D>2$ retransmissions.

Without any retransmissions (i.e., immediately after the new data packet of $N$
bits was received), the conditional probability density function (PDF) of the
received bit reliability $\rb$ can be written as,
\begin{eqnarray*}
  \ff{0} &=& \frac{1}{2} \sqrt{\frac{N_o}{\pi}} \eee^{-\frac{(\bar{r}N_o -2
      \sqrt{E_b})^2}{4 N_o}}   \\
  \fs{0} &=& \frac{1}{2} \sqrt{\frac{N_o}{\pi}} \eee^{-\frac{(\bar{r}N_o + 2
      \sqrt{E_b})^2}{4 N_o}}.
\end{eqnarray*}

Since the bits are selected for retransmission based on their reliability, we
first obtain the BER conditioned on the reliability interval,
$L_D \leq |\rb|\leq U_D$, for some positive constants $U_D \geq L_D\geq 0$.
This BER is equal to the probability of the error event `$\eee$' that $S_1$ was
transmitted, however, $S_2$ is decided at the receiver when
$- U_D\leq \rb \leq - L_D$. With no retransmissions, the probability of such
error event is equal to,
\begin{equation*}
  \begin{split}
    \Pr{\eee|S_1,D=0} = \Pr{- U_0\leq \rb \leq - L_0 \,\big| S_1} \\
    = \int_{-U_0 \sqrt{E_b}}^{- L_0 \sqrt{E_b}} \! \ff{0} \, \df \bar{r} \\
    = \Qfun{\sqrt{2 \gammab}\left(\frac{L_0}{2\gammab}+1\right)}- 
    \Qfun{\sqrt{2 \gammab} \left(\frac{U_0}{2\gammab}+1 \right)}
  \end{split}
\end{equation*}
where $\Qfun{x}=\int_x^\infty \frac{1}{\sqrt{2\pi}}\eee^{-t^2/2}\df t$ is the
Q-function, and due to symmetry, we have, $\Pr{e|S_1,D=0}= \Pr{e|S_2,D=0}$.
Assuming the a priori transmission probabilities, $\Pr{S_1}=\Pr{S_2}=1/2$, the
overall average BER for $L_0=0$ becomes,
\begin{equation}\label{eq:11}
  \BER{0}= \Qfun{\sqrt{2 \gammab}}- \Qfun{\sqrt{2 \gammab}
    \left(\frac{U_0}{2\gammab}+1\right)}.
\end{equation}
Note that, for $U_0\rightarrow\infty$, the BER \eref{eq:11} corresponds to the
BER of uncoded binary antipodal signaling, i.e., $\BERr=\Qfun{\sqrt{2\gammab}}$
\cite{Proakis}. Subsequently, the probability that the reliability of the
received bit is in the interval $L_D \leq |\rb| \leq U_D$ without any
retransmissions is computed as,
\begin{eqnarray*}
  \PRr{0} &=& \Pr{L_0 \leq |\rb|\leq U_0 \,\Big|S_1}\\
  &=& \int_{{- U_0}}^{- L_0} \! \ff{0}  \, \df\rb +
  \int_{{ L_0}}^{ U_0} \! \ff{0} \, \df\rb \\
  &=& \Qfun{\sqrt{2 \gammab}\left(\frac{L_0}{2\gammab}+1\right)} - 
  \Qfun{\sqrt{2 \gammab} \left(\frac{U_0}{2\gammab}+1\right)}+ \\
  && \Qfun{\sqrt{2 \gammab}\left(\frac{L_0}{2\gammab}-1\right)} -  
  \Qfun{\sqrt{2 \gammab} \left(\frac{U_0}{2\gammab}-1\right)}.
\end{eqnarray*}
For $L_0=0$, this probability is equal to,
\begin{equation*}
  \PRr{0}= 1\!-\! \Qfun{\sqrt{2 \gammab}
    \left(\frac{U_0}{2\gammab}\!+\!1\right)}\!-\!  
  \Qfun{\sqrt{2 \gammab} \left(\frac{U_0}{2\gammab}\!-\!1\right)}
\end{equation*}
and $\PRr{0}\rightarrow 1$ for $U_0 \rightarrow\infty$. Due to symmetry, and
for equally probable symbols $S_i$, we get, $\PR{0}=\PRr{0}=\PRrr{0}$.
Furthermore, if $Z$ is the (random) number of bits having the reliabilities in
the interval, $0 \leq |\rb| \leq U_0$, the mean value of $Z$,
$\E{Z}=N\,\PR{0}$, can be used to optimize the retransmission window size. For
example, by letting $W=\E{Z}$, we can control the target BER after the first
retransmission.

In general, it is convenient to re-scale the received bits after every
retransmission and subsequent MRC combining to maintain the constant average
energy of the received symbols. In particular, the received symbol after
combining the initially received sample $\rb_0$ with the retransmitted and
scaled samples $\rb_d$, $d=1,2,\ldots,D$, can be written as,
\begin{equation*}
  \rb_{\MRC} = \frac{1}{D+1} \left(\rb_0 + \sum_{d=1}^{D} {\rb}_d \right).
\end{equation*} 

In the sequel, let $L_d=0$ for $\forall d\geq 0$, and, without loss of
generality, we assume that the packet of $N$ symbols $S_1$ was transmitted.

\subsection{BER with one retransmission}

After the first transmission of the packet of $N$ bits, only those bits that
have the reliabilities in the interval $0\leq |\rb|\leq U_0$ are requested to
be retransmitted. After the first retransmission, the bit samples are combined
as,
\begin{equation*}
  \rb_{\MRC} = \left\{\begin{array}{cl}
      \frac{1}{2} (\rb_0+\rb_1) &\mbox{for}\quad |\rb_0| \leq U_0\\
      \rb_0 &\mbox{for}\quad |\rb_0| > U_0.
    \end{array} \right.
\end{equation*}
The random variable $\rb_0$ has the PDF, $\ff{0}$, whereas the PDF of the
random variable $\rb_1$ can be obtained by conditioning on $|\rb_0| \leq U_0$.
Since the random received samples $\rb_0$ and $\rb_1$ are statistically
independent, the PDF of $\rb_{\MRC}$ is given by the convolution (denoted as
$\ast$), i.e.,
\begin{equation*}
  \fMM{1}{\MRC} = \frac{\fff{0} \phi_1(\rb,U_0)}{ \Pr{|\rb_0|\leq U_0\,\big|S_1,
      D=0}} \ast \fff{0}  
\end{equation*} 
where
\begin{equation*}
  \phi_{1}(\rb,U_0) = \eta(\rb+ U_0)\left(1- \eta(\rb- U_0)\right) 
\end{equation*} 
and $\eta(\rb)$ is the unit-step function, i.e., $\eta(\rb)=1$ if $\rb\geq 0$,
and $0$ otherwise. After some manipulations, the PDF of $\rb_{\MRC}$ can be
written as,
\begin{equation*}
  \fMM{1}{\MRC} = \frac{\chi_1(\rb,U_0)}{ \Pr{|\rb_0|\leq U_0\,\big|S_1,D=0}}
\end{equation*} 
where we defined the function,
\begin{eqnarray*}
  \chi_d (\rb,U_0)&=& \frac{1}{4} \sqrt{\frac{(d+1) N_o}{\pi}}
  \eee^{-\frac{\left(\rb-2 \sqrt{\frac{\gammab}{N_o}}\right)^2}{\frac{4}{(d+1)
        N_o}}}\times\\ &&
  \Big\{\erf{\sqrt{\frac{(d+1) N_o}{4 d}}(U_0 - \rb)}+\\ &&
    \erf{\sqrt{\frac{(d+1) N_o}{4 d}}(U_0 + \rb)}\Big\}
\end{eqnarray*}    
and $\erf{\cdot}$ denotes the error function \cite{Proakis}.
    
Assuming that all bits with the reliabilities $|\rb_0|\leq U_0$ are
retransmitted, and then combined using the MRC, the overall $\BER{1}$ after the
first retransmission is computed as \cite{Proakis},
\begin{eqnarray*}
  \BER{1}&=& \Pr{e|d=0,S_1} \Pr{d=0|S_1} +\\ && \Pr{e|d=1,S_1} \Pr{d=1 |S_1}\\
  &=& \int_{-\infty}^{-U_0} \! \fff{0} \, \df\rb +
  \int_{-\infty}^{0} \! \chi_1(\rb,U_0) \, \df\rb.
\end{eqnarray*}
In the appendix, we show that $\BER{1}$ can be accurately approximated as,
\begin{eqnarray*}
  \BER{1} &\approx& \Qfun{\sqrt{2 \gammab} (\frac{U_0}{2\gammab}
                    +1)}+\Qfun{\sqrt{4 \gammab}} - \\ &&
      \sum_{k=1}^2 \frac{A_k}{\sqrt{1+2 B_k}} \Big\{\eee^{-
      \funmin{1}{U_0} \gammab} \Qfun{\argplus{1}{U_0} \sqrt{\gammab}}+ \\&& 
    \qquad\qquad \eee^{-\funplus{1}{U_0} \gammab} \Qfun{\argminus{1}{U_0} 
     \sqrt{\gammab}}\Big\}
\end{eqnarray*}
where the coefficients $A_k$ and $B_k$, are given in the appendix, and the
auxiliary functions,
\begin{eqnarray*}
  \funmin{d}{U}&=&\frac{B_k (d+1)(2 - U/\gammab)^2 }{2 d(1+\frac{2 B_k}{d})}\\
  \funplus{d}{U}&=&\frac{B_k (d+1)(2 + U/\gammab)^2 }{2 d(1+\frac{2 B_k}{d})}
\end{eqnarray*}
\begin{equation*}
  \argminus{d}{U} = \frac{1 - \frac{B_k U }{d\gammab}}{\sqrt{\frac{1+\frac{2
          B_k}{d}}{2(d+1)}}},\qquad
  \argplus{d}{U} = \frac{1 + \frac{B_k U}{d\gammab}}{\sqrt{\frac{1+\frac{2
          B_k}{d}}{2(d+1)}}}.
\end{equation*}

\subsection{BER with two retransmissions}

Consider now the case with two retransmissions. After the second
retransmission, the received bits having the reliabilities $|\rb| \leq U_1$ are
combined with the retransmitted samples $\rb_2$. The threshold value $U_1$ is
chosen to be greater than $U_0$ due to improvement of the bit reliabilities
after the first retransmission. More generally, we assume that,
\begin{equation*}
  U_{D-1} \geq U_{D-2} \ldots \geq U_1 \geq U_0.
\end{equation*}
Hence, the bit-samples after two retransmissions can be written as,
\begin{equation}\label{eq:13}
  \rb_{\MRC} = \left\{%
    \begin{array}{cl} \frac{1}{3} ({\rb}_0+{\rb}_1+{\rb}_2) &
    \mbox{for}\quad |\rb_0| \leq U_0,\ \frac{1}{2}|{\rb}_0+{\rb}_1|\leq U_1 \\ 
      \frac{1}{2} ({\rb}_0+{\rb}_1) & \mbox{for} \quad |\rb_0| \leq U_0,\
                                      \frac{1}{2}|{\rb}_0+{\rb}_1|> U_1 \\ 
      \frac{1}{2} ({\rb}_0+{\rb}_2) & \mbox{for} \quad U_0 < |\rb_0| \leq U_1\\ 
      \rb_0 & \mbox{for} \quad |\rb_0|>U_1.\\
    \end{array} \right.
\end{equation}
However, in order to make the analysis mathematically tractable, we merge the
first two conditions in \eref{eq:13}, and consider instead the retransmission
scheme,
\begin{equation}\label{eq:13a}
  \rb_{\MRC} = \left\{%
    \begin{array}{cl} \frac{1}{3} ({\rb}_0+{\rb}_1+{\rb}_2) &
    \mbox{for}\quad |\rb_0| \leq U_0\\
      \frac{1}{2} ({\rb}_0+{\rb}_2) & \mbox{for} \quad U_0 < |\rb_0| \leq U_1\\ 
      \rb_0 & \mbox{for} \quad |\rb_0|>U_1\\
    \end{array} \right.
\end{equation}
having the upper-bound performance than the scheme \eref{eq:13}. The scheme
\eref{eq:13a} can be interpreted as making the decision about the number of
retransmissions for each bit already after receiving the initial data packet of
$N$ bits. Even though the scheme \eref{eq:13a} may unnecessarily retransmit
some bits even if their reliability have already reached the desired threshold,
our numerical results indicate that, for $D=2$, the performance difference is
not significant (less than $1$ dB). Consequently, the number of retransmissions
for each bit can be determined by quantization of the initial reliabilities
$|{\rb}_0|$ using the thresholds $U_0$ and $U_1$. The bits having the
reliabilities $|\rb_0| \leq U_0$ are always combined with two other
retransmitted bits, whereas the received bits with the reliabilities
$ U_0<|\rb_0| \leq U_1$ are combined with exactly one retransmission.

The PDF $\fM{2}{\MRC}$ of $\rb_{\MRC}$, for $|\rb|\leq U_0$, is given as,
\begin{equation*}
  \fM{2}{\MRC} = \frac{\chi_1 (\rb,U_0)}{\Pr{|\rb_0| \leq U_0\,\big|S_1,d=0}}
\end{equation*} 
and for $U_0 < |\rb_0| \leq U_1$, it is given as,
\begin{equation*}
  \fM{2}{\MRC}=\frac{\lambda_1(\rb,U_1,U_0)}{\Pr{U_0<|\rb_0|\leq
      U_1\,\big|S_1,d=0}}
\end{equation*} 
where
\begin{equation*}
  \begin{array}{l}
    \lambda_d (\rb,U_d,U_{d-1})= \frac{1}{4}
    \sqrt{\frac{(d+1) N_o }{\pi}}\ \exp\!\left(-\frac{\left(\rb-2
          \sqrt{\frac{\gammab}{N_o}}\right)^2}{\frac{4}{(d+1) N_o}}\right)
    \times\\\quad \Bigg\{\erf{\sqrt{\frac{(d+1)N_o}{4d}}(U_d+\rb)} +
    \erf{\sqrt{\frac{(d+1) N_o}{4d}}(U_d -\rb)}-\\\qquad\erf{\sqrt{\frac{(d+1)
          N_o}{4 d}}(U_{d-1}+\rb)}\!-\!\erf{\sqrt{\frac{(d+1) N_o}{4d}}(U_{d-1}
      -\rb)}\!\!\Bigg\}.
  \end{array}
\end{equation*}
For $|\rb_0|>U_1$, the PDF $\fM{2}{\MRC}$ is given by the PDF $\fff{0}$. The
overall $\BER{2}$ is obtained using the law of the total probability, i.e.,
\begin{eqnarray*}
  \BER{2}&=& \sum_{d=0}^2 \Pr{e|d,S_1} \Pr{d | S_1}
  = \int_{-\infty}^{-U_1} \! \ff{0} \, \df\rb\\ &&
  \qquad + \int_{-\infty}^{0} \! \lambda_1 (\rb,U_1,U_0) \, \df\rb +
  \int_{-\infty}^{0} \! \chi_2(\rb,U_0) \, \df\rb
\end{eqnarray*}
where the probability of the second retransmission of the same bit is,
$\Pr{d=2|S_1}= \Pr{\rb \leq -U_1\,\big|S_1,d=1}$. As for one retransmission,
$\BER{2}$ can be approximated (see the appendix),
\begin{equation*}
  \begin{array}{l}
  \BER{2}\approx \Qfun{\sqrt{2 \gammab}(\frac{U_1}{2\gammab}+1)}+
  \Qfun{\sqrt{6\gammab}}\\ \qquad
   -\sum_{k=1}^2 \frac{A_k}{\sqrt{1+B_k}} \Big\{\eee^{- 
      \funmin{2}{U_0} \gammab} \Qfun{\argplus{2}{U_0} \sqrt{\gammab}} \\
      \qquad\qquad + \eee^{-
      \funplus{2}{U_0} \gammab} \Qfun{\argminus{2}{U_0} \sqrt{\gammab}}\Big\}\\
  \qquad+ \sum_{t=1}^2 \frac{A_k}{\sqrt{1+ 2 B_k}} \Big\{\eee^{- \funmin{1}{U_0}
    \gammab} \Qfun{\argplus{1}{U_0} \sqrt{\gammab}} \\ 
     \qquad + \eee^{- \funplus{1}{U_0} \gammab} \Qfun{\argminus{1}{U_0} 
       \sqrt{\gammab}}\\\qquad - \eee^{- \funmin{1}{U_1}
    \gammab}  \Qfun{\argplus{1}{U_1} \sqrt{\gammab}}\\ 
    \qquad -\eee^{- \funplus{1}{U_1}
    \gammab} \Qfun{\argminus{1}{U_1} \sqrt{\gammab}}\Big\}.
  \end{array}
\end{equation*}

The probability that the received reliability is in the interval
$U_0\leq |\rb|\leq U_1$ is evaluated as,
\begin{equation*}
  \begin{array}{l}
  \PR{1}= \Pr{|\rb|\leq U_1\,\big|S_1,d=0} \Pr{d=0|S_1}+\\\ \Pr{|\rb|\leq
    U_1\,\big|S_1,d=1} \Pr{d=1 |S_1} = \int_{-U_1}^{U_1} \! \ff{0} \,
    \df\rb\\\qquad\qquad- \int_{U_0}^{U_0}\!\ff{0}\,\df\rb + 
    \int_{-U_1}^{U_1} \!\chi_1(\rb,U_0)\,\df\rb.
  \end{array}
\end{equation*}
This probability can be efficiently approximation as,
\begin{equation*}
  \begin{array}{l}
  \PR{1} \approx \Qfun{\sqrt{2 \gammab} (\frac{U_0}{2\gammab}+1)}+
           \Qfun{\sqrt{2\gammab} (\frac{U_0}{2\gammab}-1)}\\\qquad
           -\Qfun{\sqrt{2 \gammab}
    (\frac{U_1}{2\gammab}+1)}-\Qfun{\sqrt{2\gammab} (\frac{U_1}{2\gammab}-1)}\\
  \\\qquad+ \Qfun{2\sqrt{\gammab}-U_1/\sqrt{\gammab}}-\Qfun{2\sqrt{\gammab}+
    U_1/\sqrt{\gammab}} \\\qquad-\sum_{k=1}^2 \frac{A_k}{2 \sqrt{1+2 B_k}}
  \Big\{\eee^{- \funplus{1}{U_0} \gammab} \big\{\erf{\arggplusm{1}{U_0}
    \sqrt{\gammab}} \\\qquad +\erf{\arggminp{1}{U_0} \sqrt{\gammab}}\big\} 
    \!+\! \eee^{- \funmin{1}{U_0} \gammab } \big\{\erf{\arggplusp{1}{U_0}
    \sqrt{\gammab}}\\\qquad+ 
    \Sign{\T{U_0}} \erf{\arggminm{1}{U_0} \sqrt{\gammab}} \big\}\Big\}
  \end{array}
\end{equation*}  
where the auxiliary functions,
\begin{eqnarray*}
  \arggplusm{d}{U}&=& \frac{\frac{U_D}{2\sqrt{\gammab}}+\frac{1}{\sqrt{N_o}} +
    \frac{B_k(U_D-U)}{d\sqrt{\gammab}}}{\frac{1+\frac{2 B_k}{d}}{(d+1) N_o}}\\
  \arggminp{d}{U}&=& \frac{\frac{U_D}{2\sqrt{\gammab}} - \frac{1}{\sqrt{N_o}} +
    \frac{B_k(U_D + U)}{d\sqrt{\gammab}}}{\frac{1+\frac{2 B_k}{d}}{(d+1) N_o}}
\end{eqnarray*}
\begin{eqnarray*}
  \arggplusp{d}{U}&=&\frac{\frac{U_D}{2\sqrt{\gammab}}+\frac{1}{\sqrt{N_o}}+
    \frac{B_k(U_D+U)}{d\sqrt{\gammab}}}{\frac{1+\frac{2 B_k}{d}}{(d+1) N_o}}\\
  \arggminm{d}{U}&=&\frac{\left|\T{U}\right|}{2}\sqrt{(d+1) N_o(1+\frac{2
      B_k}{d})}\\
  \T{U}& =& \frac{U_D}{\sqrt{\gammab}} - \frac{\frac{2}{\sqrt{N_o}}+\frac{2 B_k
      U}{\sqrt{\gammab}}}{1+2 B_k}
\end{eqnarray*}
and $\Sign{\cdot}$ is the sign function.

\subsection{BER with multiple retransmissions}

For $D\geq 1$ retransmissions, the overall BER is calculated as,
\begin{equation*}
  \BER{D}= \sum_{d=0}^D \Pr{e|d,S_1} \Pr{d | S_1}.
\end{equation*}
Recall that, for the sake of mathematical tractability, we assume that the
received bits with their initial reliability within the interval
$|\rb| \leq U_0$ will be retransmitted $D$ times, the bits with the reliability
in the interval $U_0 < |\rb|\leq U_1$ will be retransmitted $D-1$ times and so
on. The number of the received copies for each reliability range after $D$
retransmissions is shown in \fref{Fig:1}. Consequently, the overall BER can be
expressed using the functions $\chi_D(\rb,U_{D-1})$ and
$\lambda_D(\rb,U_D ,U_{D-1})$ as,
\begin{equation*}
  \begin{array}{l}
  \BER{D}=  \int_{-\infty}^{-U_{D-1}} \! \ff{0} \, \df\rb \\\qquad\quad +
  \sum_{i=1}^{D-1} \int_{-\infty}^{0} \! \lambda_i(\rb,U_{D-i},U_{D-i-1})
  \, \df\rb+\int_{-\infty}^{0} \! \chi_D(\rb,U_0) \, \df\rb 
  \end{array}
\end{equation*}
where the middle term is zero for $D=1$ retransmission. This BER can be
accurately approximated as,
\begin{equation*}
  \begin{array}{l}
  \BER{D} \approx \Qfun{\sqrt{2 \gammab} \left(\frac{U_{D-1}}{2\gammab}+
      1\right)}+ \Qfun{\sqrt{2 \gammab (D+1)}}\\\qquad  - \sum_{k=1}^2
  \frac{A_k}{\sqrt{1+\frac{2 B_k}{D}}} \Big\{\eee^{- \funmin{D}{U_0} \gammab }
  \Qfun{\argplus{D}{U_0} \sqrt{\gammab}} \\\qquad+\eee^{- \funplus{D}{U_0} \gammab}
  \Qfun{\argminus{D}{U_0} \sqrt{\gammab}}\Big\} \\\quad + \sum_{i=1}^{D-1}
    \sum_{k=1}^2
  \frac{A_k}{\sqrt{1+\frac{2 B_k}{i}}} \Big\{\eee^{- \funmin{i}{U_{D-i-1}}
    \gammab} \Qfun{\argplus{i}{U_{D-i-1}} \sqrt{\gammab}}\\\qquad + \eee^{-
    \funplus{i}{U_{D-i-1}} \gammab} \Qfun{\argminus{i}{U_{D-i-1}} \sqrt{\gammab}}
  \\\qquad- \eee^{- \funmin{i}{U_{D-i}} \gammab} \Qfun{\argplus{i}{U_{D-i}} 
    \sqrt{\gammab}}\\\qquad -\eee^{- \funplus{i}{U_{D-i}} \gammab}
    \Qfun{\argminus{i}{U_{D-i}} \sqrt{\gammab}}\Big\}.
  \end{array}
\end{equation*}
The probability $\PR{D}$ that the received reliability is in the interval
$U_{D-1}\leq |\rb|\leq U_D$ is evaluated as,
\begin{eqnarray*}
  \PR{D}&=&\!\! \int_{-U_{D}}^{U_{D}} \! \ff{0} \, \df\rb - \int_{-U_{0}}^{U_{0}}
  \! \ff{0} \, \df\rb \\ &&\!\!+\sum_{i=1}^{D-1} \int_{-U_{D}}^{U_{D}} \! \lambda_i
  (\rb,U_{D-i},U_{D-i-1}) \, \df\rb + \int_{-U_{D}}^{U_{D}} \! \chi_D(\rb,U_0)
  \, \df\rb  
\end{eqnarray*}
with the third term being zero for $D=1$ retransmission. It is again possible
to obtain the approximation of the probability $\PR{D}$ for $D>2$. However, the
resulting expression is much more involved than in the case of $\PR{1}$ and
$\PR{2}$, so it is not presented here.

\subsection{BER for slowly varying fading channels}

\begin{figure*}[!th]
  \centering
  \includegraphics[scale=0.65]{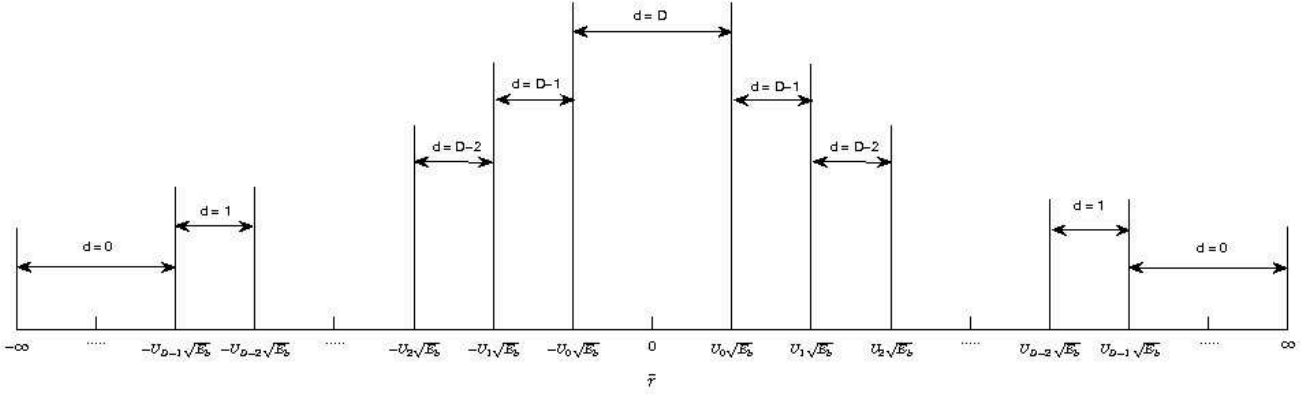}
  \caption{The quantization levels of the bit reliability with $D$
    retransmissions.}
  \label{Fig:1}
\end{figure*}

Since the wireless links are typically subject to time-varying fading, we can
average the BER expressions obtained in the previous subsection over the
time-varying received power, for example, following the chi-square
distribution. In this case, the received SNR has the PDF,
$f_{\gammab}(\gammab) = \frac{1}{\gammabr} \exp\!\left(-\frac{\gammab}{
    \gammabr}\right)$, where $\gammabr=\E{\gammab}$ is the average SNR.
Moreover, we assume that the SNR variations are sufficiently slow (e.g.,
assuming that the nodes are nomadic or even stationary), so the SNR can be
considered to be constant during the first packet transmission as well as
during the subsequent $D$ retransmissions. Then, the long-term average BER is
evaluated using the expectation \cite{Proakis},
\begin{equation*}
  \aBER{D} = \int_{0}^{\infty} \!\BERf{D}  f_{\gammab}(\gammab)\,\df\gammab.
\end{equation*}
The computationally efficient formula for calculating $\aBER{D}$ is displayed
below \fref{Fig:1} on the next page. Similarly, the average probability
$\PR{D}$ is evaluated as,
\begin{equation*}
  \PRR{D} = \int_{0}^{\infty} \!\PR{D}(\gammab) f_{\gammab}(\gammab)\,\df\gammab.
\end{equation*}

\begin{figure*}[!t]
\begin{equation*}
  \begin{array}{l}
    \aBER{D} \approx 1-\frac{1}{2} \left(\frac{4}{\gammabr(U_{D-1}/\gammab+
        2)^2} +1 \right)^{-1/2} - \frac{1}{2}\left( 1+\frac{1}{\gammabr(D+1)}
    \right)^{-1/2} \\\qquad+ \sum\limits_{k=1}^t\frac{A_k}{\sqrt{1+\frac{2
          B_k}{D}}} \Bigg\{ \left(1+\gammabr \Big(\funmin{D}{U_0} +
      \frac{\argplus{D}{U_0}}{\sqrt{2}}(\frac{\argplus{D}{U_0}}{\sqrt{2}}+
      \sqrt{\frac{1}{\gammabr} + \funmin{D}{U_0}+\frac{{\argplus{D}{U_0}}^2
        }{2}}\,\Big) \right)^{-1} \\\qquad+\left(1+\gammabr (\funplus{D}{U_0}+
      \frac{\argminus{D}{U_0}}{\sqrt{2}}(\frac{\argminus{D}{U_0}}{\sqrt{2}}+
      \sqrt{\frac{1}{\gammabr}+\funplus{D}{U_0} + \frac{{\argminus{D}{U_0}}^2}
        {2}})\right)^{-1}\Bigg\}+ \sum\limits_{i=1}^{D-1} \sum\limits_{k=1}^t
    \frac{A_k}{\sqrt{1+ \frac{2 B_k}{i}}} \times\\\ \times \Bigg\{
    \left(1+\gammabr \Big(\funmin{i}{U_{D-i-1}} + \frac{\argplus{i}{U_{D-i-1}}}
      {\sqrt{2}}(\frac{\argplus{i}{U_{D-i-1}}}
      {\sqrt{2}}+\sqrt{\frac{1}{\gammabr} + \funmin{i}{U_{D-i-1}} +
        \frac{{\argplus{i}{U_{D-i-1}}}^2}{2}}\,\Big)\right)^{-1} \\
    \qquad + \left(1+\gammabr \Big(\funplus{i}{U_{D-i-1}} +
      \frac{\argminus{i}{U_{D-i-1}}}{\sqrt{2}}(\frac{\argminus{i}{U_{D-i-1}}}
      {\sqrt{2}}+\sqrt{\frac{1}{\gammabr} + \funplus{i}{U_{D-i-1}} +
        \frac{{\argminus{i}{U_{D-i-1}}}^2}{2}}\,\Big) \right)^{-1}\\\qquad
    -  \left(1+\gammabr \Big(\funmin{i}{U_{D-i}} + \frac{\argplus{i}{U_{D-i}}}
      {\sqrt{2}}(\frac{\argplus{i}{U_{D-i}}}{\sqrt{2}}+\sqrt{\frac{1} {\gammabr}
        +\funmin{i}{U_{D-i}}+\frac{{\argplus{i}{U_{D-i}}}^2}{2}}\,\Big)
    \right)^{-1} \\\qquad - \left(1+\gammabr \Big(\funplus{i}{U_{D-i}} +
      \frac{\argminus{i}{U_{D-i}}}{\sqrt{2}}(\frac{\argminus{i}{U_{D-i}}}
      {\sqrt{2}}+\sqrt{\frac{1}{\gammabr} + \funplus{i}{U_{D-i}} +
        \frac{{\argminus{i}{U_{D-i}}}^2}{2}}\,\Big)\right)^{-1} \Bigg\}.
  \end{array}
\end{equation*}
\hrulefill
\vspace*{4pt}
\end{figure*}

\section{Bitwise Retransmission Protocols}

We consider three specific bitwise retransmission strategies and optimize their
parameters to maximize the transmission reliability rather than to maximize
their throughput. We also verify mathematical formulas derived in the previous
section by computer simulations assuming error-free feedback. This assumption
is revisited at the end of this section.

\subsection{Fixed rate technique} 

In this design, we assume a constant retransmission window size $W$ determined
as, 
\begin{equation*}
  W= \round{\frac{N}{D} \left(\frac{1}{\Rf} -1\right)}
\end{equation*} 
for a priori given parameters $N$, $D$ and the forward rate $\Rf$. Since the
window size $W$ is constrained as $1 \leq W \leq N$, the possible forward rates
are restricted to the interval, $\frac{1}{1+D} < \Rf \leq \frac{N}{D+N}$. Thus,
for $W=N$, the rate $\Rf=\frac{1}{1+D}$, and the proposed retransmission scheme
corresponds to a block repetition code (BRC) with $D$ repetitions of the
original packet. We have the following proposition.

\begin{proposition}\label{prop:1}
  For a given SNR, the fixed rate retransmission technique achieves the minimum
  BER for some specific value of the rate $\Rf$. The optimum value of $\Rf$
  minimizing the BER increases with the SNR.
\end{proposition}
\propref{prop:1} can be proved by letting the derivative
$(\df/\df\Rf)\,\BER{D}$ to be equal to zero. For large values of SNR and $N$,
the forward rate $\Rf$ approaches its maximum value of $1$, so in such case,
the overhead due to retransmissions can be neglected. Moreover, when $N$ is
large or $W=N$, the BER of the fixed rate retransmission scheme approaches the
BER of the BRC.
 
\begin{figure}[!t]
  \centering
  \psfrag{xx}[][][1.0]{$\Rf$}
  \psfrag{yy}[][][1.0]{BER}
  \includegraphics[scale=0.515]{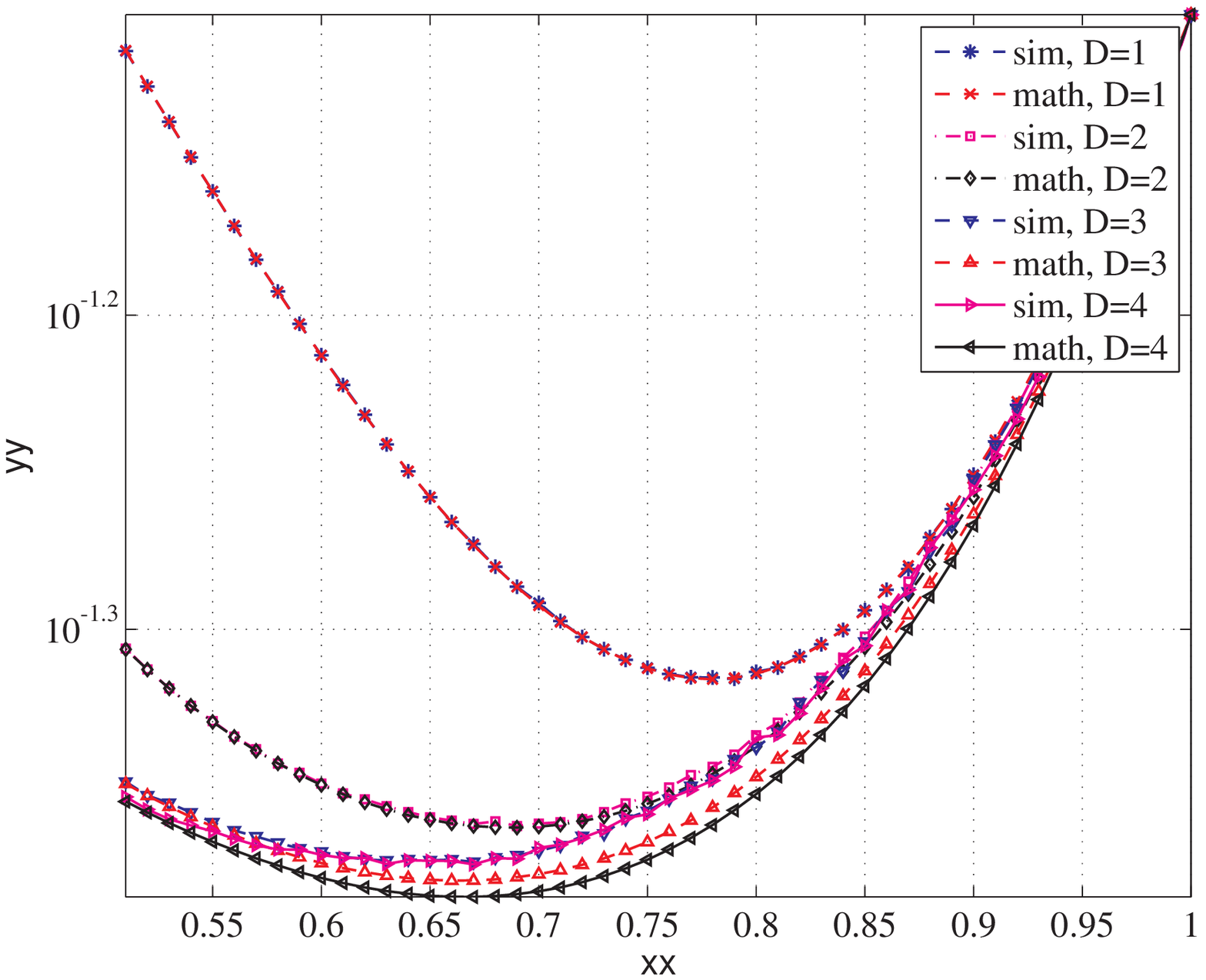}
  \caption{The BER versus the forward rate $R_f$ for $\SNR=0$ dB and the
    different number of retransmissions $D$.}
  \label{Fig:2}
\end{figure}

\begin{figure}[!t]
  \centering
  \psfrag{xx}[][][1.0]{$\Rf$}
  \psfrag{yy}[][][1.0]{BER}
  \includegraphics[scale=0.495]{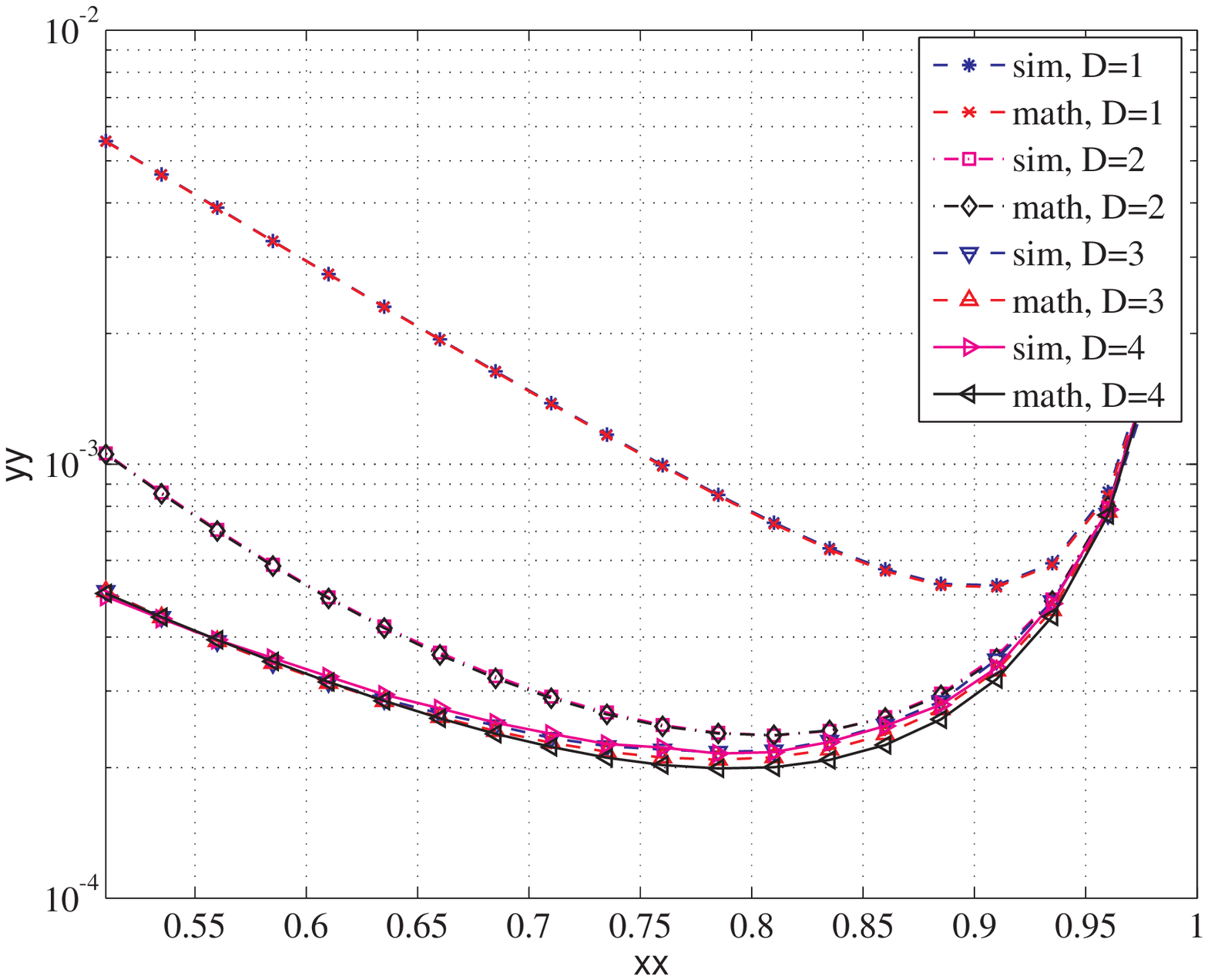}
  \caption{The BER versus the forward rate $R_f$ for $\SNR=5$ dB and the
    different number of retransmissions $D$.}
  \label{Fig:20}
\end{figure}

\fref{Fig:2} and \fref{Fig:20} show the BER of the forward link versus the
forward rate $\Rf$ for the different number of retransmissions $D$ at two SNR
values $\gammab=0$ dB and $\gammab=5$ dB, respectively. We observe that the
minimum BER value is more pronounced (i.e., the optimization is more important)
for larger values of SNR. The BER curves in \fref{Fig:2} and \fref{Fig:20} also
compare the simulations with the approximate expressions given in the previous
section; for $D>1$, the difference between the approximate expressions and the
simulations is negligible.

\begin{figure}[!t]
  \centering
  \psfrag{xx}[][][1.0]{$\gammab$ [dB]}
  \psfrag{yy}[][][1.0]{$\Rf$}
  \includegraphics[scale=0.5]{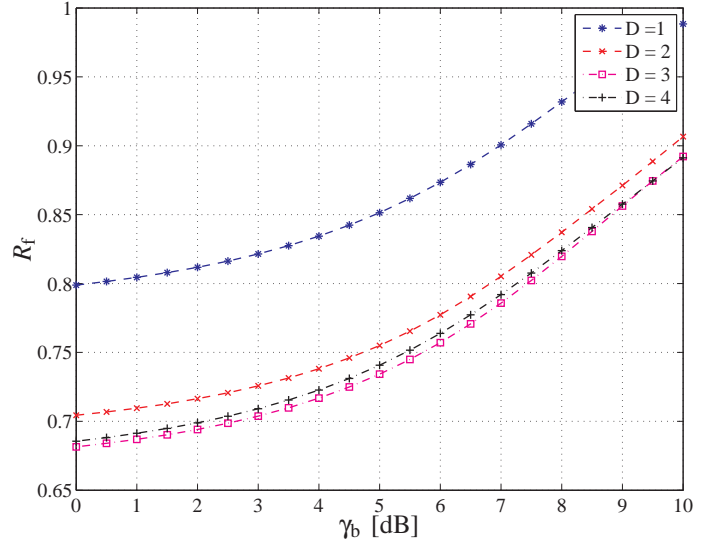}
  \caption{The rate $R_f$ yielding the minimum BER versus the SNR $\gammab$ for
    the different number of retransmissions $D$.}
  \label{Fig:3}
\end{figure}

\begin{figure}[!t]
  \centering
  \psfrag{xx}[][][1.0]{$\gammab$ [dB]}
  \psfrag{yy}[][][1.0]{$\BER{D}$}
  \includegraphics[scale=0.5]{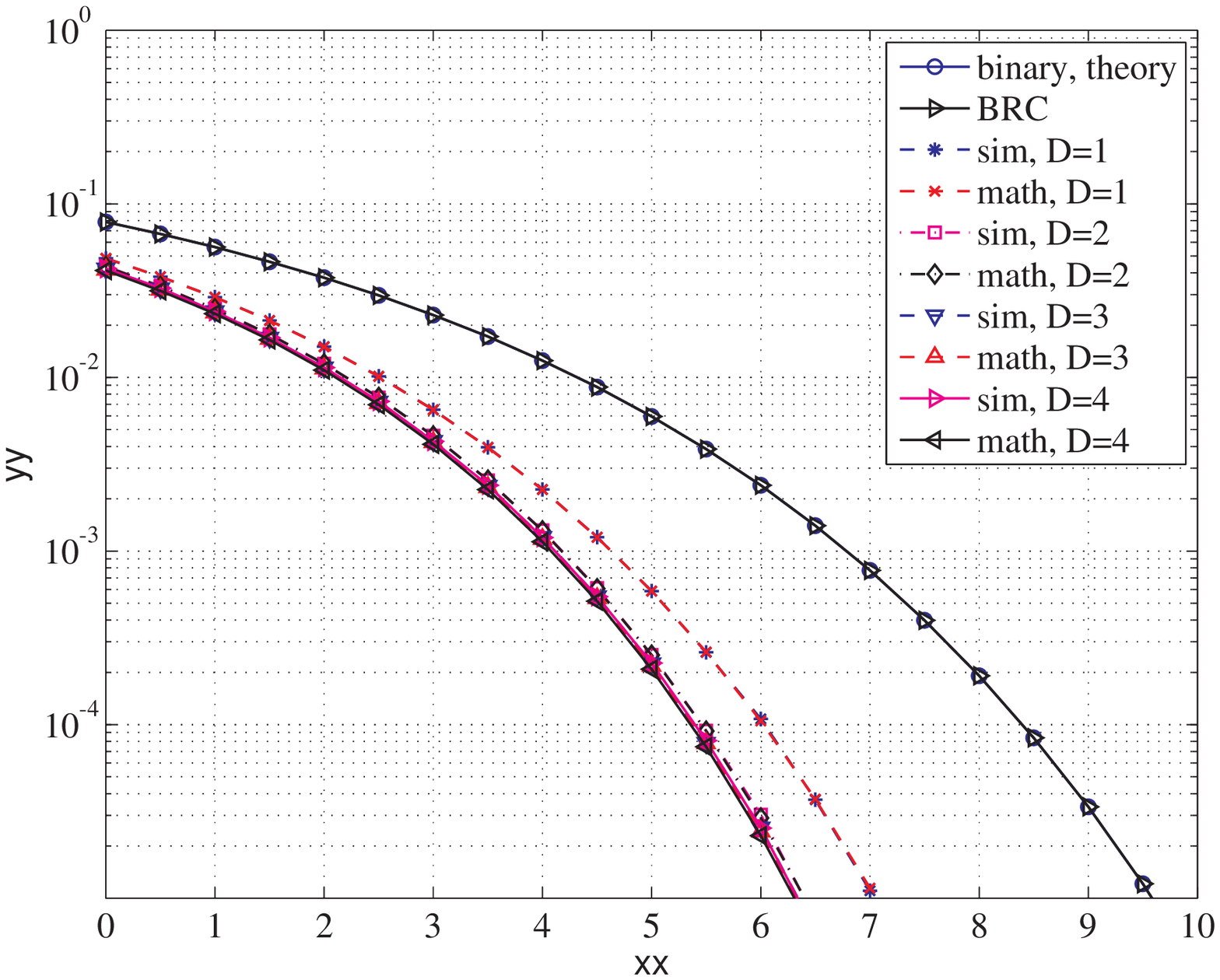}
  \caption{The $\BER{D}$ versus the SNR $\gammab$ for the different number of
    retransmissions $D$.}
  \label{Fig:4}
\end{figure}

The forward rates $\Rf$ yielding the minimum BER are shown in \fref{Fig:3} for
the different number of retransmissions $D$. Finally, \fref{Fig:4} shows the
BER versus the SNR $\gammab$ for the different number of retransmissions $D$
assuming that the forward rate $\Rf$ is optimized for each SNR value $\gammab$
to minimize the achieved BER. Note again that such optimization becomes more
effective if the SNR is increased.

\subsection{Fixed window technique}

As for the fixed rate technique, the retransmission window size is fixed, and
it is determined as,
\begin{equation*}
  W=\round{N\,P}
\end{equation*}
where the retransmission decision thresholds $U_d$ are set, so that the
probabilities $P=\PR{0} =\PR{1} = \ldots = \PR{{D-1}}$, which have been
obtained in the previous section, have the same value. Then, given $N$, $D$ and
$W$, the forward rate is calculated as,
\begin{equation*}
  \Rf = \frac{1}{1+D\,W/N}.
\end{equation*} 
We have the following proposition.
\begin{proposition}\label{prop:2}
  For a given SNR, the fixed window technique achieves the minimum BER for some
  value of the retransmission window size. The optimum window size value
  decreases with the SNR.
\end{proposition}
\propref{prop:2} can be again proved by letting the derivative,
$(\df/\df W)\,\BER{D}$ to be equal to zero. When the ratio $W/N$ approaches
unity, the BER of the fixed window technique approaches the BER of the BRC. On
the other hand, when the ratio $W/N$ approaches zero, the retransmission
overhead can be neglected.

\begin{figure}[!th]
  \centering
  \psfrag{xx}[][][1.0]{$W/N$}
  \psfrag{yy}[][][1.0]{BER}
  \includegraphics[scale=0.5]{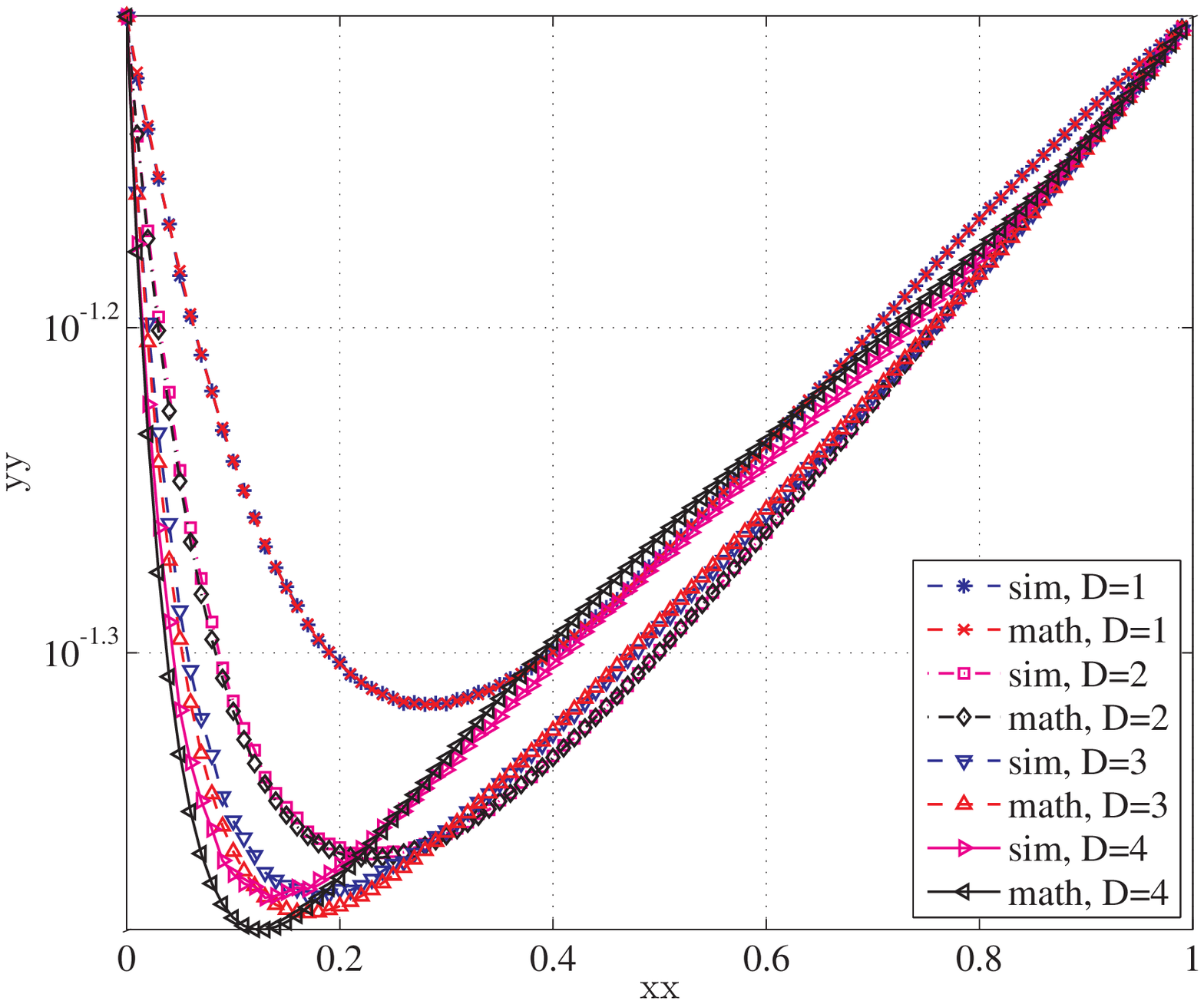}
  \caption{The BER versus the normalized window size $W/N$ for $\SNR=0$ dB and
    the different number of retransmissions $D$.}
  \label{Fig:5}
\end{figure}

\begin{figure}[!th]
  \centering
  \psfrag{xx}[][][1.0]{$W/N$}
  \psfrag{yy}[][][1.0]{BER}
  \includegraphics[scale=0.5]{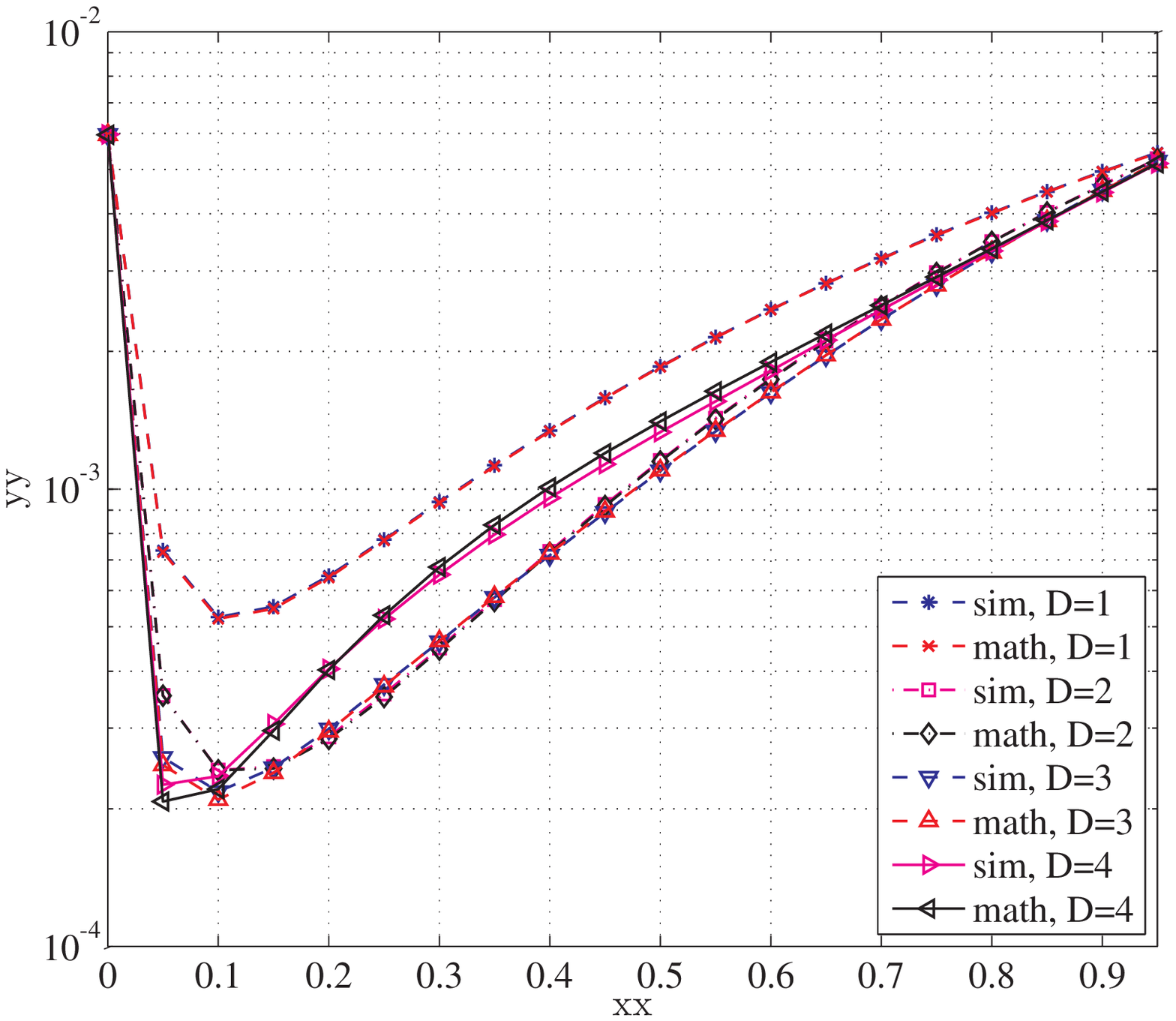}
  \caption{The BER versus the normalized window size $W/N$ for $\SNR=5$ dB and
    the different number of retransmissions $D$.}
  \label{Fig:50}
\end{figure}

\fref{Fig:5} and \fref{Fig:50} show the BER versus the normalized window size
$W/N$ for the different number of retransmissions $D$ and the SNR values
$\gammab=0$ dB and $\gammab=5$ dB, respectively. We again observe a negligible
difference between the approximate and the simulated BER curves, especially for
larger values of $D$. In addition, the minimum BER values are more pronounced
when the SNR is increased.

\begin{figure}[!th]
  \centering
  \psfrag{xx}[][][1.0]{$\gammab$ [dB]}
  \psfrag{yy}[][][1.0]{$W/N$}
  \includegraphics[scale=0.5]{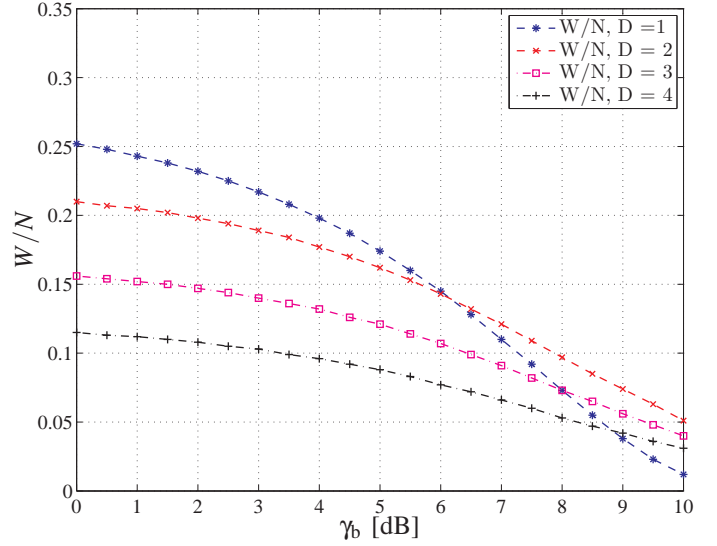}
  \caption{The normalized window size $W/N$ yielding the minimum BER versus the
    SNR $\gammab$ for the different number of retransmissions $D$.}
  \label{Fig:6}
\end{figure}

\begin{figure}[!th]
  \centering
  \psfrag{xx}[][][1.0]{$\gammab$ [dB]}
  \psfrag{yy}[][][1.0]{$\BER{D}$}
  \includegraphics[scale=0.5]{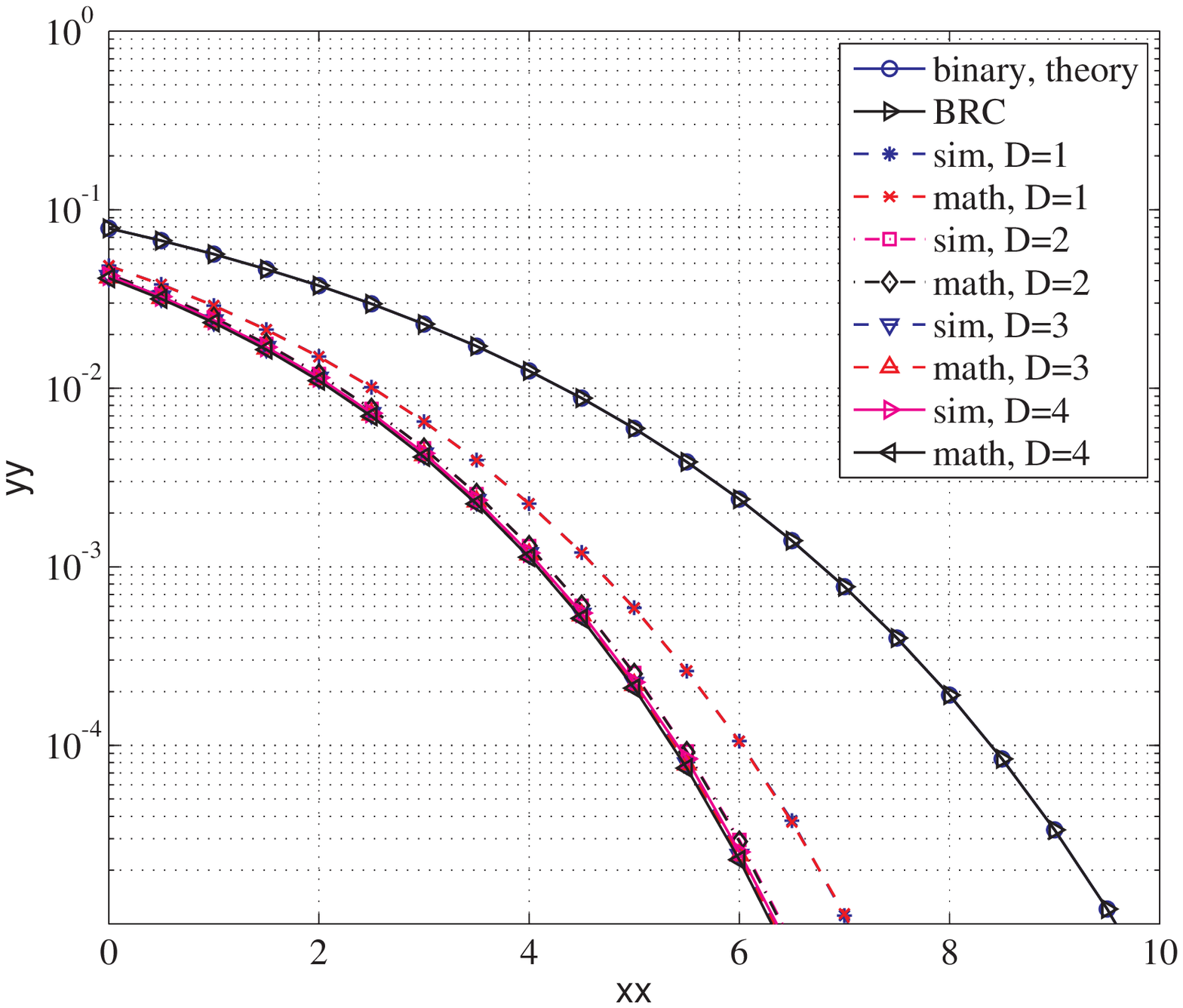}
  \caption{The $\BER{D}$ versus the SNR $\gammab$ for the different number of
    retransmissions $D$.}
  \label{Fig:7}
\end{figure}

\begin{figure}[!th]
  \centering
  \psfrag{xx}[][][1.0]{$\gammab$ [dB]}
  \psfrag{yy}[][][1.0]{BER}
  \includegraphics[scale=0.5]{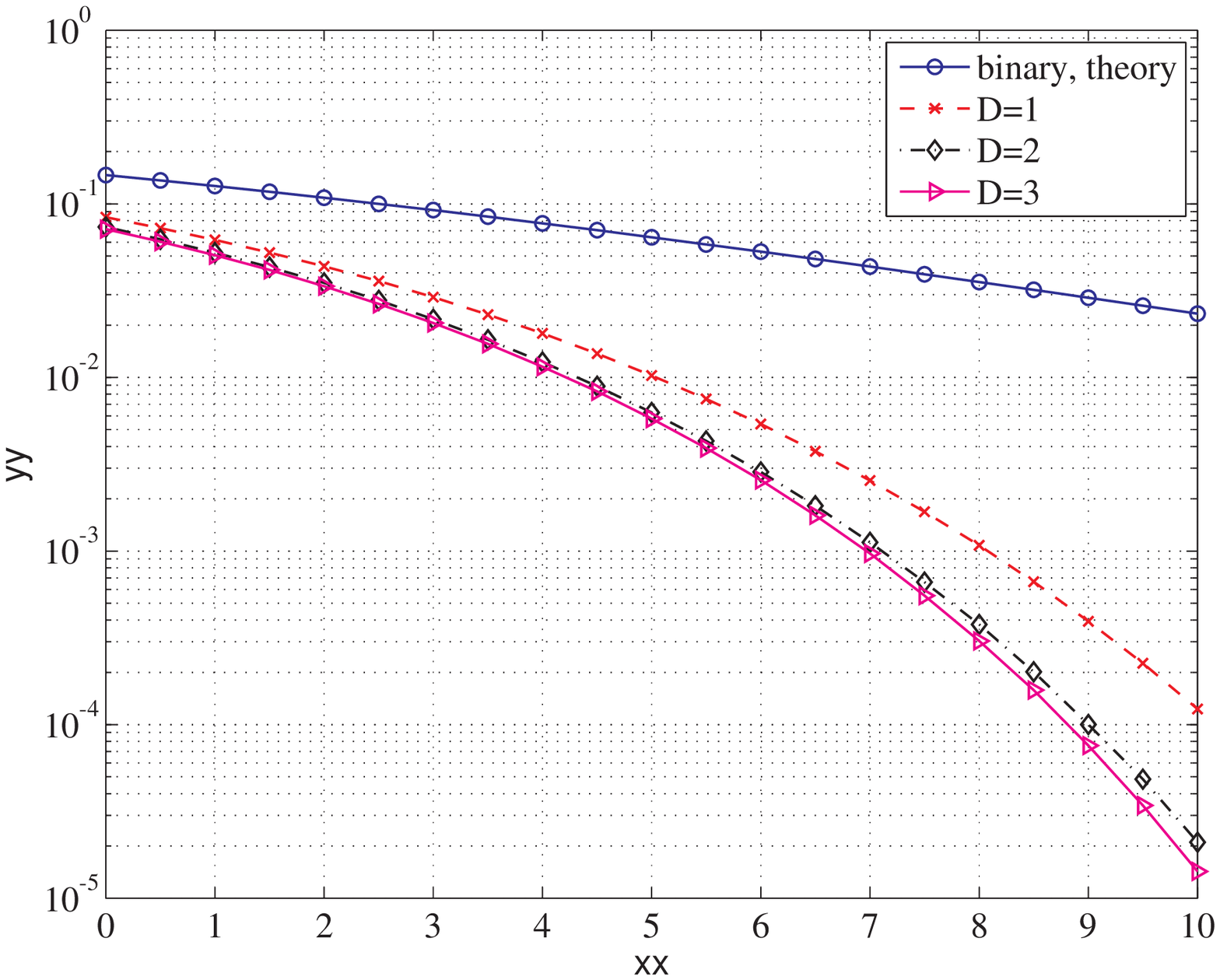}
  \caption{The average BER versus the SNR $\gammab$ over the slowly varying
    chi-square distributed transmission power.}
  \label{Fig:a7}
\end{figure}

\fref{Fig:6} shows the normalized window size $W/N$ corresponding to the
minimum BER for the different number of retransmissions $D$. \fref{Fig:7} shows
the BER versus the SNR $\gammab$ for the different number of retransmissions
$D$ assuming the optimum value of $W/N$ for each $\gammab$ minimizing the BER.
Note that such minimization of the BER is more effective when the SNR is large.
Finally, for the different number of retransmission $D$, \fref{Fig:a7} shows
the BER for a slowly varying chi-square distributed, unit-mean received power,
provided that the parameters are optimized to minimize the BER.

\subsection{Fixed threshold technique}

Unlike the previous two techniques, the fixed threshold technique allows for
different retransmission window sizes while assuming the single constant
reliability threshold, $U_0=U_1=\ldots=U_{D-1}= U$, during each retransmission.
Given $U$, we obtain the probabilities $\PR{d}$, $d=1,2,\ldots,D$, defined
previously, and calculate the retransmission window sizes as,
\begin{equation*}
  W_d = \round{N \PR{d}}.
\end{equation*}  
The corresponding forward rate is given as,
\begin{equation*}
  \Rf = \frac{1}{1+\sum_{d=1}^D W_d/N}.
\end{equation*} 
We have the following proposition.
\begin{proposition}\label{prop:3}
  For a given SNR, the fixed threshold technique achieves the minimum BER value
  for some specific threshold $U$. This optimum threshold value is increasing
  with the SNR.
\end{proposition} 
\propref{prop:3} can be again proved by letting the derivative,
$(\df/\df U)\,\BER{D}$ to be equal to zero.

\begin{figure}[!th]
  \centering
  \psfrag{xx}[][][1.0]{$U/\sqrt{E_b}$}
  \psfrag{yy}[][][1.0]{BER}
  \includegraphics[scale=0.5]{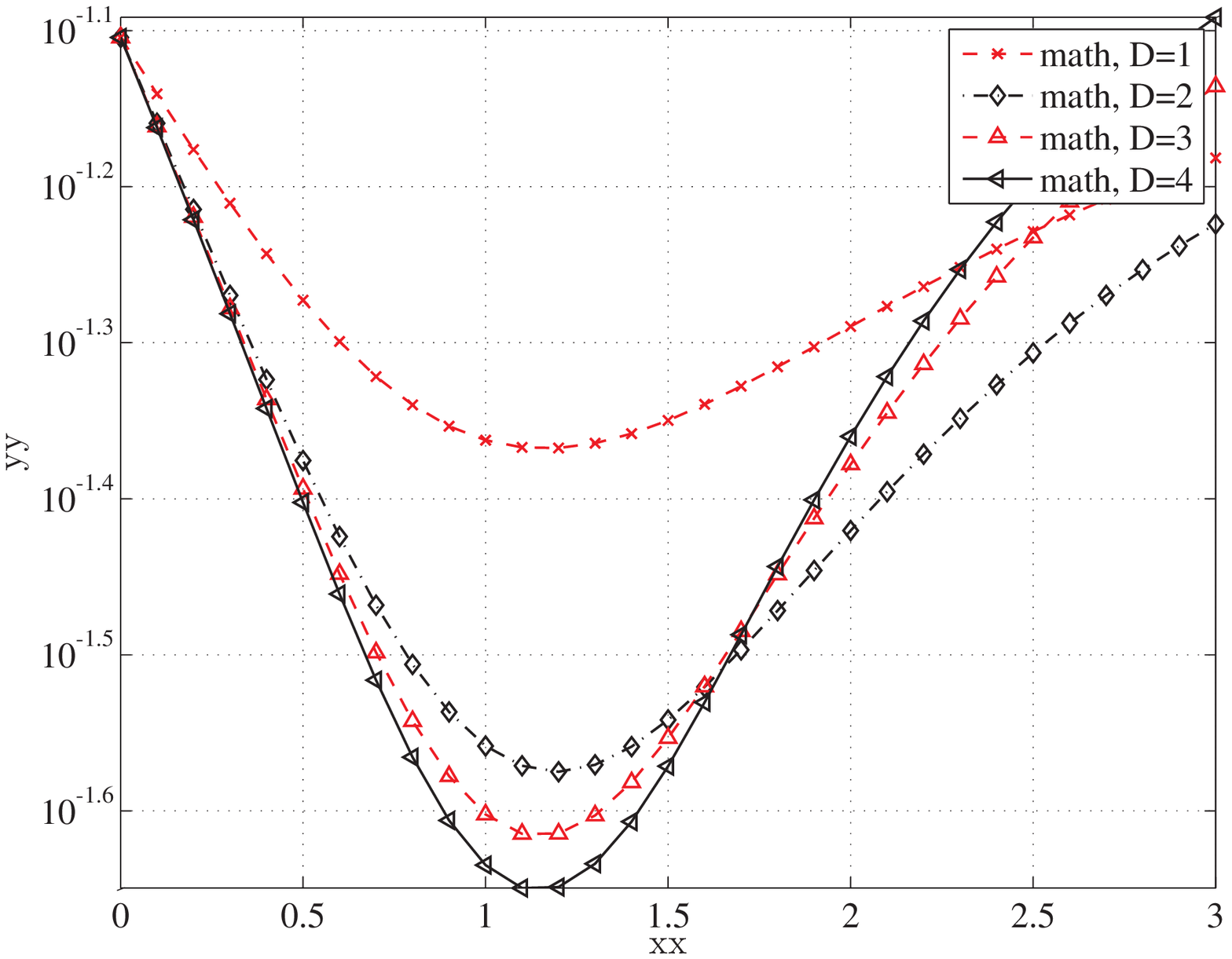}
  \caption{The BER versus the normalized threshold $U/\sqrt{E_b}$ for the SNR
    $\gammab=0$ dB and the different number of retransmissions $D$.}
  \label{Fig:8}
\end{figure}

\begin{figure}[!th]
  \centering
  \psfrag{xx}[][][1.0]{$U/\sqrt{E_b}$}
  \psfrag{yy}[][][1.0]{BER}
  \includegraphics[scale=0.5]{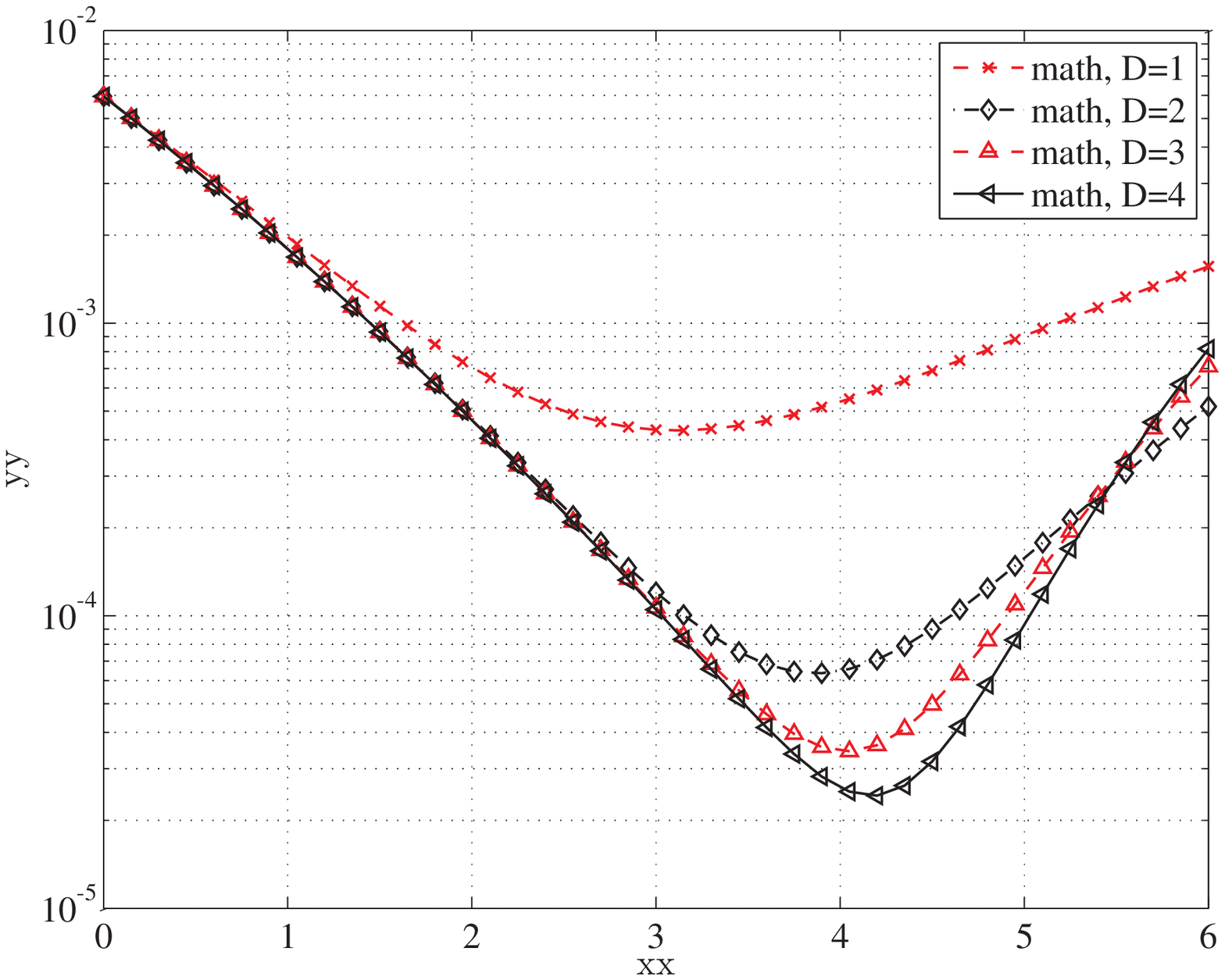}
  \caption{The BER versus the normalized threshold $U/\sqrt{E_b}$ for the SNR
    $\gammab=5$ dB and the different number of retransmissions $D$.}
  \label{Fig:80}
\end{figure}

\begin{figure}[!th]
  \centering
  \psfrag{xx}[][][1.0]{$\gammab$ [dB]}
  \psfrag{yy}[][][1.0]{$U/\sqrt{E_b}$}
  \includegraphics[scale=0.5]{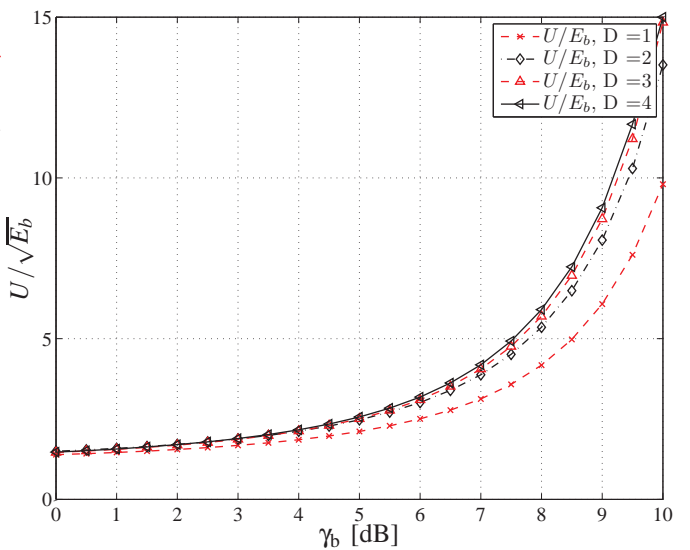}
  \caption{The values of the normalized threshold $U/\sqrt{E_b}$ corresponding
    to the minimum BER versus the SNR $\gammab$ for the different number of
    retransmissions $D$.}
  \label{Fig:9}
\end{figure}

\begin{figure}[!th]
  \centering
  \psfrag{xx}[][][1.0]{$\gammab$ [dB]}
  \psfrag{yy}[][][1.0]{$\BER{D}$}
  \includegraphics[scale=0.5]{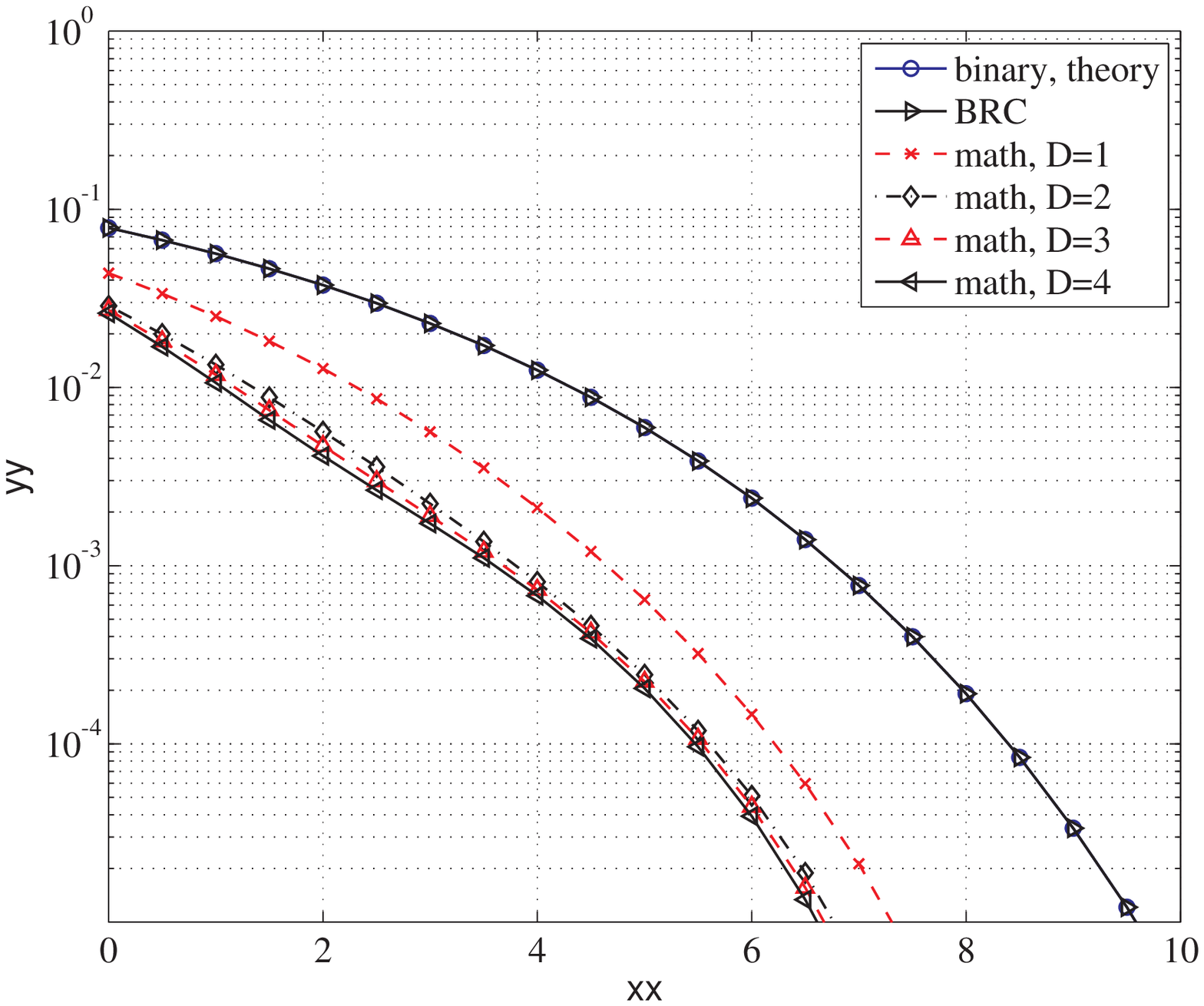}
  \caption{The $\BER{D}$ versus the SNR $\gammab$ for the different number of
    retransmissions $D$.}
  \label{Fig:10}
\end{figure}

\fref{Fig:8} and \fref{Fig:80} show the BER versus the normalized threshold
$U/\sqrt{E_b}$ for the different number of retransmissions $D$ and the SNR
values $\gammab=0$ dB and $\gammab=5$ dB, respectively. Since the approximate
BER expressions were already verified for the other two techniques considered,
the BER curves in \fref{Fig:8} and \fref{Fig:80} only show these derived
expressions. We again observe that the minimum BER values are more apparent
when the SNR is increased. \fref{Fig:9} shows the normalized thresholds
$U/\sqrt{E_b}$ having the minimum BER for the different number of
retransmissions $D$. Finally, \fref{Fig:10} shows the BER versus the SNR
$\gammab$ for different $D$ assuming the optimum thresholds $U/\sqrt{E_b}$ for
each SNR $\gammab$ that achieves the minimum BER. We again find that
minimization of the BER by optimizing the threshold $U/\sqrt{E_b}$ is more
effective when the SNR is increased.

\subsection{Feedback signaling}

There are two main issues when designing practical feedback signaling schemes.
The first issue is how to constrain the number of feedback bits. The second
issue is the transmission errors of feedback bits.

Sending a small number of feedback bits in a dedicated packet is very
inefficient due to the associated protocol overheads. In practice, it is common
to reserve a few bits within the packet payload for the feedback signaling, so
the packet overhead is shared by the feedback as well as data. In such case, it
is beneficial to minimize the number of feedback bits in order to increase the
data payload. Here, we reduce the number of feedback bits sent from the
destination to the source over the reverse link by considering a deterministic
sequence of bit permutations synchronously generated at the transmitter and at
the receiver to encode the feedback message. Such sequence is conveniently
generated as pseudo-random permutations of $N$-tuples, $(1,2,\ldots,N)$, using
the two synchronized RNGs. In addition, the transmitter and the receiver RNGs
synchronously advance by exactly $2^{C_1}$ permutations every symbol period.
For every generated permutation, the receiver checks whether a sufficient
number of bits with small reliabilities fall into a predefined window of $W$
bits; for instance, we can assume that the retransmission window is represented
by the the first $W$ bit positions. When such permutation is found, the
feedback message to notify the source is a binary representation of the
permutation order modulo $2^{C_1}$. The source then selects the corresponding
$W$ bits in the packet, and retransmits them to the destination. In particular,
for the window size $W$, the number of permutations $K$ searched until the
desired one has been found can be expressed as,
\begin{equation*}
  K=2^{C_1}I+(K\ \mathrm{mod}\, 2^{C_1})
\end{equation*}
where $I$ is an integer number of the idle symbol periods. The key property is
that,
\begin{equation*}
  (K\ \mathrm{mod}\, 2^{C_1}) \ll \binom{N}{W}
\end{equation*}
whereas, on average, $\E{K}=\binom{N}{W}$. Hence, the feedback message is only
represented by $C_1$ bits. The value of $C_1$ is a design parameter, and it
trade-offs the feedback message size, and the required delay until the start of
the next retransmission \cite{Hassanien10}.

\begin{figure}[!t]
  \centering
  \psfrag{xx}[][][1.0]{$\gammab$ [dB]}
  \psfrag{yy}[][][1.0]{$\E{K}$}
  \includegraphics[scale=0.5]{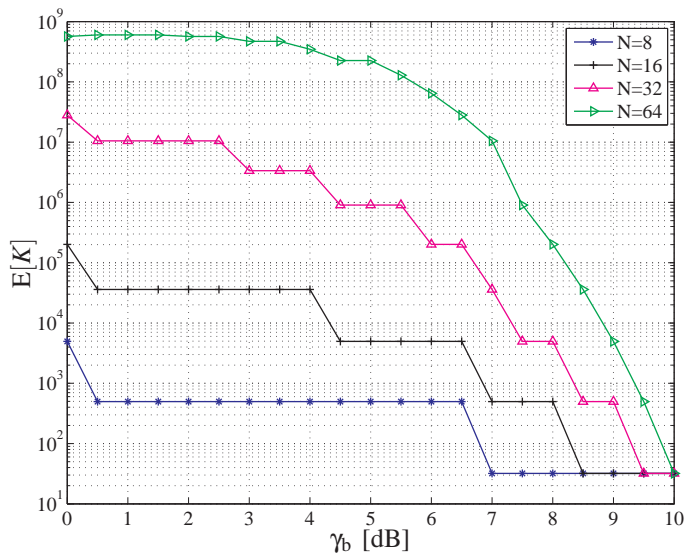}
  \caption{The expected number of permutations $\E{K}$ versus the SNR $\gammab$
    and the different packet length $N$.}
  \label{Fig:a2}
\end{figure}

More importantly, given some target BER, the window size $W$ and the average
number of searched permutations $\E{K}$ are decreasing with the SNR. As shown
in \fref{Fig:a2}, $\E{K}$ is only about $30$ for all $N\leq 64$ when the SNR is
at least $10$ dB, and thus, the feedback of $C_1=5$ bits can be used to find
the desired permutation during one symbol period.

\begin{figure}[!t]
  \centering \psfrag{xx}[][][1.0]{$\gammab$ [dB]}
  \psfrag{yy}[][][1.0]{$C_1^\ast$}
  \includegraphics[scale=0.5]{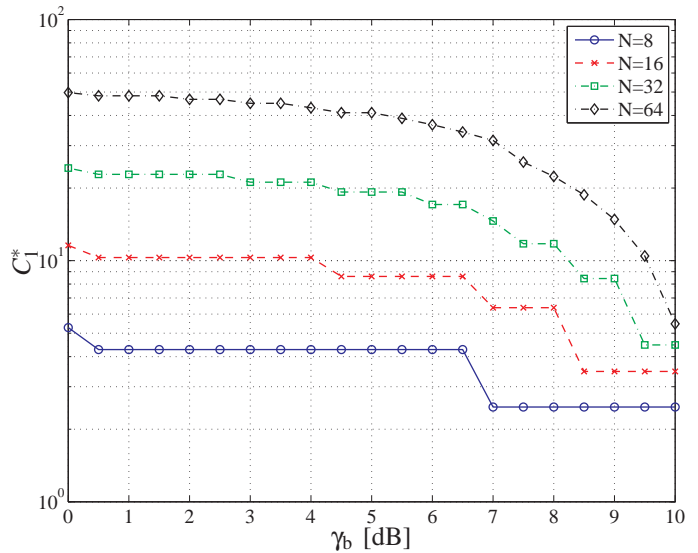}
  \caption{The optimum values $C_1^\ast$ with $D=1$ retransmission versus the
    SNR $\gammab$ and the different packet length $N$.}
  \label{Fig:a3}
\end{figure}

\begin{figure}[!t]
  \centering \psfrag{xx}[][][1.0]{$\gammab$ [dB]}
  \psfrag{yy}[][][1.0]{$\zeta_1^\ast$}
  \includegraphics[scale=0.5]{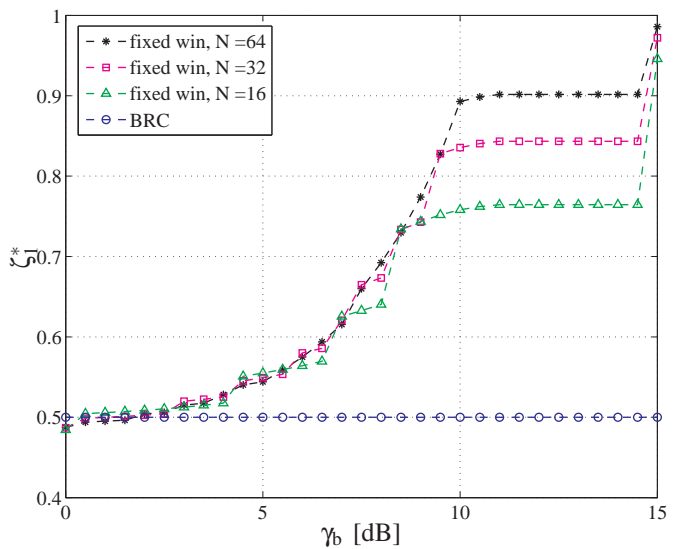}
  \caption{The throughput $\zeta_1^\ast$ corresponding to the optimum values
    $C _1^\ast$ versus the SNR $\gammab$ and the different packet length $N$.}
  \label{Fig:a5}
\end{figure}

The average forward throughput with one retransmission can be defined as,
\begin{equation*}
  \zeta_1 = \frac{N}{\E{I+1}}.
\end{equation*}
By simulations, we found that there exists an optimum value $C_1^\ast$ which
minimizes the expected delay $\E{I+1}$, i.e., which maximizes the throughput
$\zeta_1$. This optimum value is given as,
\begin{equation*}
  C_1^\ast= \round{-0.5+\log_2\E{K}}
\end{equation*}
and it is plotted in \fref{Fig:a3} for different values of the packet length
$N$. The corresponding maximum throughput $\zeta_1^\ast$ is shown in
\fref{Fig:a5}. For comparison, the throughput $1/2$ of the BRC corresponding to
the conventional stop-and-wait ARQ with one retransmission is also shown in
\fref{Fig:a5}. More importantly, we observe that the proposed scheme can
achieve better throughput than the rate $1/D$ repetition code for the same
number of retransmissions $D$, especially at the medium to large values of SNR.
Furthermore, we can show that the throughput of the fixed window technique in
the limit of very large SNR converges to,
$ \lim_{\gammab\rightarrow \infty} \zeta_1 \approx \frac{N}{N+1}$, since
$\lim_{\gammab\rightarrow \infty} W/N = 0$.

Finally, we reconsider the assumption of the error-free feedback which is often
adopted in many papers concerning the ARQ retransmission schemes. Recall that
it was shown in \cite{Makki14} that errors of 1 and 2-bit feedback messages can
be neglected if their bit error probability is less than $10^{-3}$. This result
can be readily modified for the case of multi-bit feedback messages which are
utilized in our bitwise retransmission schemes. In particular, assuming the
feedback errors are independent, and they are occurring with the probability
$p_r$, we have the following proposition.
\begin{proposition}\label{prop:4}
  The errors in the feedback messages of $C$ bits can be tolerated, provided
  that their bit-error probability is bounded as,
  \begin{equation*}
    p_r\leq 1- \Pmin^{1/C}
  \end{equation*}
  where the required minimum probability of the error-free feedback messages
  was established in \cite{Makki14} to be, $\Pmin=1-10^{-3}=99.9\%$.
\end{proposition} 
The proof of \propref{prop:4} follows from the binomial distribution of
independent and equally probable errors. Hence, the longer the feedback
message, the smaller the feedback bit error probability $p_r$ is required.

\section{Data Fusion Application}

We now illustrate the use of the proposed bitwise retransmission schemes in a
practical scenario of data fusion from a group of $L$ sensor nodes into a
single central access point (AP). A time-division multiple access (TDMA)
protocol with $(L+1)$ time slots is used to share the communication channel.
The time-division duplex (TDD) protocol further divides the available time
slots into $L$ uplink time slots to transmit data from the $L$ sensor nodes,
and one time slot is allocated for the feedback signaling from the AP. Each
time slot can carry at most $N$ bits of information including the protocol
overhead. The AP is assumed not to be battery powered, so the transmit power in
the downlink can be much larger than in the battery constrained uplink.

More specifically, we consider the following three sensor node technologies:
Zigbee 802.15.4, Wifi 802.11b and Bluetooth v. 4.2 802.15.1. The parameters of
these three technologies are summarized in \tref{tab:1}. Their BER expressions
have been obtained by fitting the sum of exponentials (the Prony method,
\cite{Loskot}) to the performance curves reported in \cite{Petrova06}. It is
obvious that Zigbee is the most energy efficient technology for the sensor
nodes, and it can operate at small SNR values.

\begin{table*}[!t]
  \centering
  \caption{Transmission parameters of the three sensor node technologies}
  \begin{tabular}{|c|c|c|c|}\hline
   \rr{2} & Zigbee & Wifi & Bluetooth \\\hline\hline
    \rr{4} BER & $P_b(\gammab)= 1.5203 \eee^{-9.5611 \gammab}$ & 
        $\begin{array}{rcl} P_b(\gammab)&=& 10.0\eee^{-3.4535\gammab} \\
           && + 1.1066\eee^{-2.0247\gammab}\end{array}$ &
        $\begin{array}{rcl} P_b(\gammab)&=& 0.2436\eee^{-0.4997\gammab} \\
           && + 0.2436\eee^{-0.4997\gammab}\end{array}$ \\
    & $\gammab$ [dB] & $\gammab$ [dB] & $\gammab$ [dB] \\
    $10^{-2}$ & -2.79 & 3.88 & 8.91 \\
    $10^{-3}$ & -1.16 & 5.43 & 10.93 \\
    $10^{-4}$ & 0.03 & 6.63 & 12.30 \\
    $10^{-5}$ & 0.96 & 7.59 & 13.34 \\
    $10^{-6}$ & 1.73 & 8.37 & 14.18 \\\hline
    \rr{2} Packet & header & preamble+header & header+CRC \\
    {[}bytes] & $6$ & $15$--$24$ & $2$+$2$ \\
    & payload & payload & payload \\
    & $127$ & $1500$ & $252$ \\
    $N$ [bits] & $(6+127)\times 8= 1064$ & $(24+1500)\times 8= 12192$ & 
    $(4+252)\times 8= 2048$ \\ \hline
    \rr{3} Segments & $14\stimes 76$, $19\stimes 56$, $28\stimes 38$ &
    $16\stimes 762$, $32\stimes381$, $48\stimes 254$ & 
    $16\stimes 128$, $32\stimes 64$, $64\stimes 32$  \\
    & $38\stimes28$, $56\stimes 19$, $76\stimes 14$ &                        
    $96\stimes 127$, $127\stimes 96$, $254\stimes 48$ &
    $128\stimes 16$, $256\stimes 8$, $512\stimes 4$ \\  
    & $133\stimes 8$, $152\stimes 7$, $266\stimes 4$ &
      $381\stimes 32$, $508\stimes 24$, $762\stimes 16$ &
      $1024\stimes 2$, $2048\stimes 1$ \\
    & $532\stimes 2$, $1064\stimes 1$ &
      $1016\stimes 12$, $1524\stimes 8$, $2032\stimes 6$ & \\
    & & $3048\stimes 4$, $4064\stimes 3$, $6096\stimes 2$ & \\ 
    & & $12192\stimes 1$ & \\\hline
  \end{tabular}
  \label{tab:1}
\end{table*}

The specific design examples of the bitwise retransmission schemes with a
constant retransmission window size employing the binomial number feedback and
the packet segmentation are listed in \tref{tab:2}. Therein, $p_f$ is the BER
of the forward link, $p_r$ is the BER of the reverse link, $\Nseg$ is the
number of segments of the disjoint partitioning of the $N$ bit packet, $\Wseg$
is the retransmission window size per segment, and $\Ctot$ is the total length
of the feedback message (in bits) required for the whole packet. Hence,
$N/\Nseg$ is the segment length, $\Wseg\Nseg=W$ is the total number of
retransmitted bits, and $\Ctot/\Nseg$ is the length of the feedback message per
segment. Furthermore, $\Pp_f$ is the probability that there are at most $\Wseg$
errors in any segment (i.e., the closer the value of $\Pp_f$ to $1.0$, the
better), and $\Pp_r$ is the probability that there is at least one error among
the $\Ctot$ feedback bits received at the source (i.e., the probability that
the feedback message is received incorrectly at the source). As shown in the
previous section, it is required that $\Pp_r<\Pmin=10^{-3}$ in order to neglect
the effect of feedback errors on the performance. Other system parameters are
given in \tref{tab:1} including the packet size $N$ and the required SNR for
given values of $p_f$.

We found that, to obtain efficient designs of the proposed bitwise
retransmission schemes, the BER $p_f$ of the forward link should be at most
$10^{-3}$, and the BER $p_r$ of the reverse link should be at most $10^{-5}$.
In terms of the minimum required SNR for bidirectional connections, the Zigbee
requires $-1.16$ dB and $0.96$ dB, the Wifi requires $5.43$ dB and $7.59$ dB,
and the Bluetooth requires $10.93$ dB and $13.34$ dB, respectively. Such SNR
levels can be satisfied for all three wireless technologies considered,
provided that the reverse link has $3$ dB larger transmit power than the
forward link. Such transmission power unbalances can be readily obtained in the
sensor networks with the centralized mains-powered AP.
 
\begin{table}[!th]
  \setlength\tabcolsep{5pt}
  \centering
  \caption{The bitwise retransmission designs with the constant window size and
    the packet segmentation}
  \begin{tabular}{|cccccccc|}\hline
    \rr{2.5} & $p_f$ & $p_r$ & $\Nseg$ & $\Wseg$ & $\Ctot$ & $\Pp_f$ & $\Pp_r$ 
    \\\hline
    \rr{3} Zigbee& $10^{-3}$ & $10^{-5}$ & $2$ & $3$ & $50$ & $0.9978$ &
    $5.0\cdot10^{-4}$ \\
    &$10^{-3}$ & $10^{-5}$ & $1$ & $4$ & $36$ & $0.9953$ & $3.6\cdot 10^{-4}$\\
    &$10^{-3}$ & $10^{-5}$ & $1$ & $5$ & $44$ & $0.9992$ & $4.4\cdot 10^{-4}$\\
    &$10^{-4}$ & $10^{-5}$ & $4$ & $1$ & $36$ & $0.9997$ & $3.6\cdot 10^{-4}$\\
    &$10^{-4}$ & $10^{-5}$ & $2$ & $1$ & $20$ & $0.9986$ & $2.0\cdot 10^{-4}$\\
    &$10^{-4}$ & $10^{-5}$ & $2$ & $2$ & $36$ & $1.0000$ & $3.6\cdot 10^{-4}$\\
    &$10^{-4}$ & $10^{-5}$ & $2$ & $3$ & $50$ & $1.0000$ & $5.0\cdot 10^{-4}$\\
    &$10^{-4}$ & $10^{-5}$ & $1$ & $1$ & $11$ & $0.9947$ & $1.1\cdot 10^{-4}$\\
    &$10^{-4}$ & $10^{-5}$ & $1$ & $2$ & $20$ & $0.9998$ & $2.0\cdot 10^{-4}$\\
    &$10^{-4}$ & $10^{-5}$ & $1$ & $3$ & $28$ & $1.0000$ & $2.8\cdot 10^{-4}$\\
    &$10^{-4}$ & $10^{-5}$ & $1$ & $4$ & $36$ & $1.0000$ & $3.6\cdot 10^{-4}$\\
    \hline
    \rr{3} Wifi& $10^{-3}$ & $10^{-6}$ & $1$ & $21$ & $220$ & $0.9928$ & 
    $2.2\cdot 10^{-4}$ \\
    & $10^{-4}$ & $10^{-6}$ & $4$ & $2$ & $92$ & $0.9962$ & $9.2\cdot 10^{-5}$\\ 
    & $10^{-4}$ & $10^{-6}$ & $3$ & $2$ & $69$ & $0.9917$ & $6.9\cdot 10^{-5}$\\
    & $10^{-4}$ & $10^{-6}$ & $2$ & $3$ & $72$ & $0.9965$ & $7.2\cdot 10^{-5}$\\
    & $10^{-4}$ & $10^{-6}$ & $2$ & $4$ & $92$ & $0.9996$ & $9.2\cdot 10^{-5}$\\
    & $10^{-4}$ & $10^{-6}$ & $1$ & $4$ & $50$ & $0.9917$ & $5.0\cdot 10^{-5}$\\
    & $10^{-4}$ & $10^{-6}$ & $1$ & $5$ & $61$ & $0.9984$ & $6.1\cdot 10^{-5}$\\
    & $10^{-4}$ & $10^{-6}$ & $1$ & $6$ & $72$ & $0.9997$ & $7.2\cdot 10^{-5}$\\
    & $10^{-4}$ & $10^{-6}$ & $1$ & $7$ & $83$ & $1.0000$ & $8.3\cdot 10^{-5}$\\
    & $10^{-4}$ & $10^{-6}$ & $1$ & $8$ & $94$ & $1.0000$ & $9.4\cdot 10^{-5}$\\
    \hline
    \rr{3} Bluetooth & $10^{-2}$ & $10^{-6}$ & $2$ & $18$ & $256$ & $0.9913$ &
    $2.6\cdot 10^{-4}$ \\
    & $10^{-3}$ & $10^{-6}$ & $2$ & $4$ & $72$ & $0.9960$ & $7.2\cdot 10^{-5}$\\
    & $10^{-3}$ & $10^{-6}$ & $1$ & $6$ & $57$ & $0.9949$ & $5.7\cdot 10^{-5}$\\
    & $10^{-3}$ & $10^{-6}$ & $1$ & $7$ & $65$ & $0.9987$ & $6.5\cdot 10^{-5}$\\
    & $10^{-3}$ & $10^{-6}$ & $1$ & $8$ & $73$ & $0.9997$ & $7.3\cdot 10^{-5}$\\
    & $10^{-4}$ & $10^{-5}$ & $4$ & $1$ & $36$ & $0.9987$ & $3.6\cdot 10^{-4}$\\
    & $10^{-4}$ & $10^{-5}$ & $2$ & $1$ & $20$ & $0.9951$ & $2.0\cdot 10^{-4}$\\
    & $10^{-4}$ & $10^{-5}$ & $2$ & $2$ & $38$ & $0.9998$ & $3.8\cdot 10^{-4}$\\
    & $10^{-4}$ & $10^{-5}$ & $1$ & $2$ & $21$ & $0.9988$ & $2.1\cdot 10^{-4}$\\
    & $10^{-4}$ & $10^{-5}$ & $1$ & $3$ & $31$ & $0.9999$ & $3.1\cdot 10^{-4}$\\
    & $10^{-4}$ & $10^{-5}$ & $1$ & $4$ & $40$ & $1.0000$ & $4.0\cdot 10^{-4}$\\
    \hline
  \end{tabular}
  \label{tab:2}
\end{table}

Next, we consider scheduling of the packet contents for our single-cell
TDMA/TDD multiple access protocol. We assume that the parameters $W$, $D$, $N$,
$L$ and $\Ctot$ are constant, even though they can be optimized for the uplink
and downlink BERs $p_f$ and $p_r$. Recall that all packets have the maximum
length of $N$ bits. In the uplink, the nodes send their data as well as
schedule the retransmitted bits for the previously transmitted data packets. In
the downlink, the AP broadcasts the retransmission requests to all sensor nodes
at once. The contents in the uplink packets are scheduled following the
following two rules.
\begin{enumerate}
\item Include the retransmitted sequences in the order corresponding to the
  previously transmitted packets. Only one retransmitted sequence per each
  previously transmitted packet can be scheduled.
\item Add data bits from the buffered information blocks to fill in the whole
  packet of $N$ bits.
\end{enumerate}
Hence, the transmitted packet can contain retransmitted bits for multiple
previously transmitted packets, and also data bits from multiple information
blocks. As an example, assuming first-in first-out (FIFO) buffering of the
information blocks of $\Nbuf=N=1064$ bits (Zigbee protocol) with $\Nseg=2$
segments, $\Wseg=2$ bits, i.e., the retransmission window of $W=2\times 2=4$
bits, $D=3$ retransmissions, and in total $\Lpac=10$ data packets to be
transmitted, the packets content schedule is shown in \tref{tab:3}. Therein, we
use the notation D$_l(n)$ to denote a sequence of $n$ bits belonging to the
$l$-th information block, and R$_{l,d}(m)$ is the sequence of $m$ retransmitted
bits in the $d$-th retransmission for the $l$-th information block where the
sequence indexes, $1\leq l\leq \Lpac$, $1\leq n\leq N$, $1\leq d\leq D$ and
$1\leq m\leq W$. We have the following proposition.

\begin{table}[!t]
   \centering
  \caption{The uplink packet contents for $N=1064$, $W=4$, $D=3$, $\Lpac=10$}
  \begin{tabular}{|c|l|}\hline
    $\#$ & packet content\\\hline\hline
    $1$ &  D$_{1}$(1064) \\\hline
    $2$ &  R$_{1,1}$(4), D$_{2}$(1060) \\\hline
    $3$ &  R$_{1,2}$(4), D$_{2}$(4), D$_{3}$(1056) \\\hline
    $4$ &  R$_{1,3}$(4), R$_{2,1}$(4), D$_{3}$(8), D$_{4}$(1048) \\\hline
    $5$ &  R$_{2,2}$(4), R$_{3,1}$(4), D$_{4}$(16), D$_{5}$(1040) \\\hline
    $6$ &  R$_{2,3}$(4), R$_{3,2}$(4), R$_{4,1}$(4), D$_{5}$(24), D$_{6}$(1028)
    \\\hline 
    $7$ &  R$_{3,3}$(4), R$_{4,2}$(4), R$_{5,1}$(4), D$_{6}$(36), D$_{7}$(1016)
    \\\hline 
    $8$ &  R$_{4,3}$(4), R$_{5,2}$(4), R$_{6,1}$(4), D$_{7}$(48), D$_{8}$(1004)
    \\\hline 
    $9$ &  R$_{5,3}$(4), R$_{6,2}$(4), R$_{7,1}$(4), D$_{8}$(60), D$_{9}$(992)
    \\\hline 
    $10$ & R$_{6,3}$(4), R$_{7,2}$(4), R$_{8,1}$(4), D$_{9}$(72), D$_{10}$(980)
    \\\hline 
    $11$ & R$_{7,3}$(4), R$_{8,2}$(4), R$_{9,1}$(4), D$_{10}$(84) \\\hline
    $12$ & R$_{8,3}$(4), R$_{9,2}$(4), R$_{10,1}$(4) \\\hline
    $13$ & R$_{9,3}$(4), R$_{10,2}$(4) \\\hline
    $14$ & R$_{10,3}$(4) \\\hline
  \end{tabular}
  \label{tab:3}
\end{table}

\begin{proposition}\label{prop:5}
  For $N\gg D\,W$, $\Nbuf=N$ and $\Lpac\geq 1$, only the very first information
  block is completely transmitted in the first packet whereas all other
  information blocks are split into exactly 2 subsequent packets. The $D$
  retransmissions are scheduled into $D$ subsequent packets immediately after
  the transmission of the corresponding information block was completed.
  Moreover, only the first $\Lpac$ transmitted packets are fully occupied with
  $N$ bits. The available transport capacity in the last $(D+1)$ transmitted
  packets can be used to transmit additional information blocks with the
  progressively shorter block lengths and/or smaller number of retransmissions.
\end{proposition}
The main assumption required in \propref{prop:5} is that the total number of
retransmitted bits $D\cdot W$ is much smaller than the block length $N$. If
this condition is not satisfied, the packet structure of the retransmission
protocol is less predictable.

We conclude our discussion about the bitwise retransmission scheme for the
uplink data fusion from $L$ sensor nodes by considering the packets structure
in the downlink. According to \propref{prop:5}, the AP (the data fusion center)
sends the retransmission requests for the $l$-th information block during the
time slots $\{1,2,\ldots,D\}$, if $l=1$, and $\{l+1,l+2,\ldots,l+D\}$, if
$l\geq 2$. However, since the information blocks with the index $l\geq 2$ are
transmitted exactly in 2 subsequent packets with the indexes $l$ and $l+1$, the
number of retransmission requests contained in the downlink packet first raises
to the maximum value of $D$ requests per sensor node. The number of the
requests then remain constant until it is gradually decremented to 1 request in
the last $(D-1)$ transmissions. Consequently, we have the last proposition.
\begin{proposition}\label{prop:6}
  The maximum number of sensor nodes $\Lmax$ which can be supported by the
  proposed retransmission scheme is bounded as,
  \begin{equation*}
    \Lmax\leq \round{\frac{N-\Novh}{D\cdot \Ctot}}
  \end{equation*}
  where $\Novh$ is the protocol overhead, and $\Ctot$ is the total number of
  feedback bits per retransmission and sensor node.
\end{proposition}
For the example presented in \tref{tab:3}, assuming $\Novh=106$ bits ($10\%$ of
$N$), we have $\Ctot=36$ bits, so $\Lmax\leq 8$. A larger number of sensor
nodes can be supported by trading off the number of feedback bits $\Ctot$ with
the number of retransmissions. In addition, it is possible that different
sensor nodes set their retransmission parameters differently, for example, to
match the SNR they experience in the uplink and downlink. In this case, the
value of $\Lmax$ in \propref{prop:6} can be calculated by assuming the maximum
total number of feedback bits, $\max(D\cdot\Ctot)$, allowed per any sensor
node.

\section{Conclusions}

A novel bitwise retransmission scheme was presented to selectively retransmit
only the bits which were received with a small reliability. The bitwise
retransmission decisions as well as combining can be done either immediately
after demodulating the received symbols, or after the channel decoding. In case
of the turbo decoding, the bit reliabilities are available as the soft-input
(i.e., channel output) or the soft-output values \cite{Big05}. The locations of
the bits to be retransmitted are reported to the source as a binomial
combination number. Since the proposed scheme does not involve any complex
operations, it does not limit the processing throughput at the receiver nor at
the transmitter.

The analysis presented in the paper assumes uncoded binary modulation for
mathematical tractability and clarity of the presentation. However, the
proposed scheme can be readily generalized to non-binary modulations by
appropriately scaling the demodulated bits. We derived the overall BER
conditioned on the error-free feedback link. The accurate closed-form BER
expressions as well as their computationally efficient approximations were
presented assuming one and two retransmissions. The approximations were
verified by computer simulations. We showed that the bitwise retransmissions
can be optimized for the given SNR, especially in the forward link, since the
BER curves are convex in the transmission rates as well as in the reliability
threshold. In addition to minimizing the BER as investigated in this paper, it
is possible to maximize the throughput instead.

We next compared the BER and throughput performances of the three specific
retransmission strategies which are referred to as the fixed rate technique,
the fixed window technique, and the fixed threshold technique. It was shown
that, for the same number of retransmissions, and the same packet length, the
proposed schemes always outperform the repetition diversity, and, in some cases
including the transmissions over time-varying channels, the performance
improvement can be significant. Furthermore, in order to reduce the number of
feedback bits, we proposed to use two synchronized RNGs at the transmitter and
at the receiver which can greatly compress the transmitted binary feedback
sequences at the expense of larger delays. We also derived a condition when the
impact of feedback errors to the overall performance can be neglected.

We then considered practical design issues of multi-user bitwise retransmission
schemes for data fusion applications where the sensor nodes forward data
packets in the uplink into a centralized AP. The retransmission requests are
broadcasted to the sensor nodes from the AP in the downlink. Assuming TDMA/TDD
and the Zigbee, Wifi and Bluetooth protocols, we presented the design examples
of retransmission parameters. These examples suggest that the efficient designs
of the bitwise retransmission schemes can be obtained provided that the BER of
the forward link is below $10^{-3}$ and the BER of the reverse link is below
$10^{-5}$. If the bidirectional link has the same or similar BERs in both
directions, the smaller BER in the reverse link can be readily achieved by
increasing the transmit power in the downlink (i.e., at the AP). We also
devised scheduling of the information and retransmission bits utilizing the
FIFO buffering of the information blocks in order to fully fill up the
transmitted packets. Finally, we calculated the upper bound on the number of
sensor nodes which can be supported by the proposed bitwise retransmission
schemes.

\appendices

\section*{Appendix}

We obtain several approximations to efficiently evaluate the integrals in the
BER expressions. The approximations are based on the Prony approximation of the
$\Qfun{x}$ function \cite{Loskot},
\begin{equation*}
  \Qfun{x}= \sum_{k=1}^2 A_k \eee^{-B_k x^2}
\end{equation*}
where $A_1=0.208$, $A_2=0.147$, $B_1=0.971$, and $B_2=0.525$.

Since $\erf{x}= 1-2\Qfun{\sqrt{2} x}$, assuming positive constants $H>0$, and
$h_i>0$, $i=1,2,3,4,5$, the expressions for $\BER{D}$ and $\PR{D}$ can be
approximated using the following expressions. These approximations are used and
verified numerically in the figures presented in Section IV.

\begin{equation*}
  \begin{array}{l}
    \int_{-\infty}^{0} \! h_1 \eee^{-\frac{(\rb-h_2)^2}{h_3}}
    \Qfun{h_4(h_5-\rb)} \, \df\rb \\ \qquad\approx \sum_{k=1}^t
    \int_{-\infty}^{0} \! h_1 \eee^{-\frac{(\rb-h_2)^2}{h_3}} A_k \eee^{-
      B_k (h_4(h_5-\rb))^2} \, \df\rb\\\qquad
    \approx \frac{h_1 A_k}{2 \sqrt{\frac{1}{h_3}+{h_4}^2 B_k}} \eee^{-\frac{B_k
        {h_4}^2 (h_2-h_5)^2}{1+h_3 {h_4}^2 B_k}} \sqrt{\pi}\,\erfc{\frac{h_2+h_3
        {h_4}^2 h_5 B_k}{\sqrt{h_3(1+h_3 {h_4}^2 B_k)}}}
    \end{array}
\end{equation*}
\begin{equation*}
  \begin{array}{l}
    \int_{-\infty}^{0} \! h_1 \eee^{-\frac{(\rb-h_2)^2}{h_3}}
    \Qfun{h_4(h_5+\rb)} \, \df\rb \\ \qquad \approx \sum_{k=1}^t
    \int_{-\infty}^{0} \! h_1 \eee^{-\frac{(\rb-h_2)^2}{h_3}} A_k \eee^{-
      B_k (h_4(h_5+\rb))^2}\,\df\rb\\ \qquad
    \approx \frac{h_1 h_3 A_k}{2 \sqrt{h_3(1+h_3 {h_4}^2 B_k)}}
    \eee^{-\frac{B_k {h_4}^2 (h_2+h_5)^2}{1+h_3 {h_4}^2 B_k}} \sqrt{\pi}\,
    \erfc{\frac{h_2-h_3 {h_4}^2 h_5 B_k}{\sqrt{h_3(1+h_3 {h_4}^2 B_k)}}}
  \end{array} 
\end{equation*}  
\begin{equation*}
  \begin{array}{l}
    \int_{-H}^{H} \! h_1 \eee^{-\frac{(\rb-h_2)^2}{h_3}} \Qfun{h_4(h_5-\rb)} \,
    \df\rb \\ \qquad \approx \sum_{k=1}^t \int_{-H}^{H} \! h_1
    \eee^{-\frac{(\rb-h_2)^2}{h_3}} A_k \eee^{- B_k (h_4(h_5-\rb))^2} \,\df\rb\\
    \qquad \approx \frac{h_1 A_k \sqrt{h_3}\sqrt{\pi}}{2 \sqrt{1+h_3 {h_4}^2
    B_k}} \eee^{-\frac{B_k {h_4}^2 (h_2-h_5)^2}{1+h_3 {h_4}^2 B_k}} \Big\{
    \erf{\frac{h_2+H+h_3 {h_4}^2 (h_5+H) B_k}{\sqrt{h_3(1+h_3 {h_4}^2 B_k)}}}\\
    \qquad +\Sign{H- \frac{h_2+h_3 {h_4}^2 h_5 B_k}{1+ h_3 {h_4}^2 B_k}}\\
    \qquad\quad \times \erf{\sqrt{\frac{1}{h_3}+{h_4}^2 B_k} \left|H-
    \frac{h_2+h_3 {h_4}^2 h_5 B_k}{1+ h_3 {h_4}^2 B_k}\right|}\Big\}
  \end{array} 
\end{equation*}
\begin{equation*}
  \begin{array}{l}
    \int_{-H}^{H} \! h_1 \eee^{-\frac{(\rb-h_2)^2}{h_3}} \Qfun{h_4(h_5+\rb)} \,
    \df\rb \\ \qquad \approx \sum_{k=1}^t \int_{-H}^{H} \! h_1
    \eee^{-\frac{(\rb-h_2)^2}{h_3}} A_k \eee^{- B_k (h_4(h_5+\rb))^2} \, \df\rb\\
    \qquad \approx \frac{h_1 h_3 A_k \sqrt{\pi}}{2 \sqrt{h_3(1+h_3 {h_4}^2
    B_k)}} \eee^{-\frac{B_k {h_4}^2 (h_2+h_5)^2}{1+h_3 {h_4}^2 B_k }}\\\qquad\times
    \Big\{\erf{\frac{H+h_2+h_3 {h_4}^2 (H- h_5) B_k}{\sqrt{h_3(1+h_3 {h_4}^2
          B_k)}}}+\erf{\frac{H-h_2+h_3 {h_4}^2 (H+h_5) B_k}{\sqrt{h_3(1+h_3
          {h_4}^2 B_k)}}}\Big\}
  \end{array} 
\end{equation*}

\end{document}